\documentclass[modern]{aastex63}

\usepackage{graphicx}
\usepackage{epstopdf}
\usepackage{ulem} 
\usepackage{color} 
\usepackage{lineno}
\usepackage{amsmath}
\usepackage{lipsum} 
\usepackage{stix} 
\usepackage{soul,xcolor} 

\linenumbers

\setstcolor{red} 

\def\lsim{\;\raise0.3ex\hbox{$<$\kern-0.75em\raise-1.1ex\hbox{$\sim$}}\;}
\def\gsim{\;\raise0.3ex\hbox{$>$\kern-0.75em\raise-1.1ex\hbox{$\sim$}}\;}

\newcommand{\xx}[1]{\!\times\!10^{#1}}
\newcommand{\epse}{\varepsilon_{e}}
\newcommand{\epsB}{\varepsilon_{B}}

\newcommand{\fNT}{f_\mathrm{NT}}
\newcommand{\fNTcrit}{f_{\mathrm{NT},\mathrm{crit}}}
\newcommand{\tobs}{t_\mathrm{obs}}

\definecolor{purple}{RGB}{200,100,255} 

\definecolor{DCW}{rgb}{1,0,0}
\definecolor{BA}{rgb}{1,0.6,0 }
\definecolor{BMV}{rgb}{0,0.0,0.99}
\definecolor{MGD}{rgb}{0.6,0,0.6}
\definecolor{Hiro}{rgb}{0,0.5,1}

\renewcommand{\baselinestretch}{1}
\def\aap{Astronomy and Astrophysics}

\def\apjl{The Astrophysical Journal Letters}

\newcount\listnorom
\newcommand\listromanDE{\global\advance \listnorom by 1
{\lowercase\expandafter{(\romannumeral\listnorom)}\ }}

\newcount\listnumber
\newcommand\listDE{\global\advance \listnumber by 1
{\lowercase\expandafter{(\number\listnumber)}\ }}

\newcount\Inum
\newcount\IInum
\newcount\IIInum
\newcount\IVnum
\def\I{\global\multiply\IInum by 0 \global\multiply\IIInum by 0
            \global\multiply\IVnum by 0 \global\advance \Inum by 1
            {\the\Inum. }}
\def\II{\global\multiply\IIInum by 0\global\multiply\IVnum by 0
       \global\advance \IInum by 1 {\the\Inum.\the\IInum. }}
\def\III{\global\multiply\IVnum by 0\global\advance \IIInum by 1
            {\the\Inum.\the\IInum.\the\IIInum. }}
\def\IV{\global\advance \IVnum by 1
            {\the\IVnum. }}
\shorttitle{Thermal electrons, SSC}
\shortauthors{Warren et al.}

\begin{document}

\title{A semi-analytic afterglow with thermal electrons and synchrotron self-Compton emission} 

\vskip24pt

\newcommand{\iTHEMS}{RIKEN Interdisciplinary Theoretical and Mathematical Sciences Program (iTHEMS), Wak\={o}, Saitama, 351-0198 Japan}
\newcommand{\ABBL}{RIKEN Cluster for Pioneering Research (CPR), Astrophysical Big Bang Laboratory (ABBL), Wak\={o}, Saitama, 351-0198 Japan}

\author[0000-0002-3222-9059]{Donald~C. Warren}%
\affiliation{\iTHEMS}
\author[0000-0003-4442-8546]{Maria Dainotti}
\affiliation{National Astronomical Observatory of Japan, NAOJ, Mitaka, Tokyo, Japan}
\affiliation{The Graduate University for Advanced Studies, SOKENDAI, Shonankokusaimura, Hayama, Miura District, Kanagawa 240-0193, Japan}
\affiliation{Space Science Institute, 4750 Walnut St, Suite 205, Boulder, CO, 80301, USA}
\author[0000-0002-0960-5407]{Maxim~V. Barkov}%
\affiliation{Institute of Astronomy, Russian Academy of Sciences, Moscow, 119017 Russia}
\author[0000-0003-4000-8341]{Bj\"orn Ahlgren}
\affiliation{\iTHEMS}
\affiliation{Department of Physics, KTH Royal Institute of Technology, and The Oskar Klein Centre, SE-10691 Stockholm, Sweden}
\author[0000-0002-2974-763X]{Hirotaka Ito}
\affiliation{\ABBL}
\author[0000-0002-7025-284X]{Shigehiro Nagataki}
\affiliation{\iTHEMS}
\affiliation{\ABBL}

\begin{abstract}

We extend previous work on gamma-ray burst (GRB) afterglows involving hot thermal electrons at the base of a shock-accelerated tail.  Using a physically-motivated electron distribution based on first-principles simulations, we compute broadband emission from radio to TeV gamma-rays.  For the first time, we present the effects of a thermal distribution of electrons on synchrotron self-Compton (SSC) emission.  The presence of thermal electrons causes temporal and spectral structure across the entire observable afterglow, which is substantively different from models that assume a pure power-law distribution for the electrons.  We show that early-time TeV emission is enhanced by more than an order of magnitude for our fiducial parameters, with a time-varying spectral index that does not occur for a pure power law of electrons.  We further show that the X-ray ``closure relations'' take a very different, also time-dependent, form when thermal electrons are present; the shape traced out by the X-ray afterglows is a qualitative match to observations of the traditional decay phase.

\end{abstract}

\section{Introduction}
\label{sec:intro}

The afterglows of gamma-ray bursts (GRBs) were initially used to localize the hosts, determine redshifts, and conclusively demonstrate the cosmological origin of these events \citep{Costa_etal_1997, vanParadijs_etal_1997, Metzger_etal_1997, BDKF1998}.  Even before the first afterglows were observed, however, it was known that broadband observations would allow for the determination of the properties of GRBs and their progenitor systems \citep[e.g.,][]{PaczynskiRhoads1993}.  A standard picture of GRB afterglows was developed, which assumed a relativistic shock accelerating electrons into a power-law distribution \citep[][and countless others]{SPN1998,GranotSari2002}; these electrons then produced, by synchrotron radiation, the photons detected at Earth.  Despite the simplicity of the model, it has been fitted with great success to numerous afterglows \citep{Perley_etal_2014, Laskar_etal_2016, Troja_etal_2019}.

The standard picture of synchrotron afterglows is not without problems, however.  Fairly early on it was realized that not all electrons were necessarily part of the power law producing the radiation; in fact, the synchrotron model is degenerate with respect to the fraction $\fNT$ of non-thermal radiating electrons \citep{EichlerWaxman2005}.  While certain observations could break the degeneracy \citep{TIN2008}, most authors assume $\fNT = 1$ on the principle of parsimony.  A second issue with the standard afterglow is how to characterize the electrons that are not part of the power-law tail.  \citet{EichlerWaxman2005} assumed that they formed a cool, thermal distribution at energies far below the base of the power law.  First-principles simulations of shock formation and particle acceleration later showed that (at least for some part of the parameter space relevant to GRBs) relativistic shocks produce a hot thermal distribution that is smoothly connected to the traditional nonthermal population \citep{SironiSpitkovsky2011,SSA2013}.

Various authors have described the effects of a hot thermal electron population on afterglow emission.  \citet{GianniosSpitkovsky2009} computed light curves and spectral evolution for a variety of mixed (thermal and nonthermal) distributions, but they used an oversimplified model of emission.  In a pair of papers, \citet{WEBN2017,Warren_etal_2018} considered the nonlinear interaction between relativistic shocks and the particle distributions they produce, as well as the consequences for GRB afterglows.  However, their model for computing emission was also simplified.  \citet{ResslerLaskar2017} produced the most thorough analysis to date of thermal electrons in GRB afterglows, solving the radiative transfer equation for both synchrotron emission and absorption within the surface of equal arrival time.  Of these studies of thermal electrons, only \citet{WEBN2017} discussed the impact of thermal electrons on TeV emission from GRB afterglows.  This was a largely theoretical concern until GRB~180720B and GRB~190114C, both of which produced photons above 300~GeV \citep{MAGIC2019Natur575, Abdalla_etal_2019Natur575}.

The standard synchrotron afterglow breaks the photon spectrum into a (possibly smoothly connected) broken power law \citep{SPN1998, GranotSari2002}.  It is therefore possible to describe the temporal ($F_{\nu} \propto t^{-\alpha}$) and spectral ($F_{\nu} \propto \nu^{-\beta}$) behavior for any desired frequency band or observer time.  One can then compute the so-called ``closure relations'', which relate $\alpha$ and $\beta$ for the afterglow.  Numerous works have computed these closure relations under a wide variety of physical conditions \citep[wind-like or constant-density ambient medium, slow- or fast-cooling electrons, presence or absence of a reverse shock, etc.][]{SPN1998, ZhangMeszaros2004, Racusin_etal_2009, Gao_etal_2013, Ryan_etal_2020}.  Other authors have applied closure relations to populations of GRBs in order to interpret observations of the standard afterglow decay phase to pinpoint a statistically favored GRB scenario \citep{Srinivasaragavan_etal_2020}; or applied the same procedure to the shallower plateau emission phase \citep{Dainotti_etal_2021PASJ}.  Closure relations have additionally been used in an attempt to identify the afterglow environment of select peculiar GRBs \citep[such as those with a plateau in both their GeV and X-ray light curve: ][]{Danotti_etal_2021ApJS255}.

The aim of this paper is to extend prior work on thermal electrons in GRB afterglows, especially their relevance to observable TeV emission.  In Section~\ref{sec:model} we outline a semi-analytical description of a GRB afterglow with hot thermal electrons.  In Section~\ref{sec:fNT} we define the two main models used in this work, as well as discussing the mostly-free parameter $\fNT$.  We solve the radiative transfer equation and present broadband spectra in Section~\ref{sec:full_dist}.  In Section~\ref{sec:LCs_indices} we discuss the effects of thermal electrons on light curves, spectral and temporal indices.  We then interpret the joint behavior of the indices, as applied to X-ray closure relations, in Section~\ref{sec:closure_rels}.  We conclude in Section~\ref{sec:conclusions}.

\section{Model}
\label{sec:model}

The model presented here may be conceptually divided into four parts: (1) a description of the hydrodynamic properties of the forward shock responsible for the afterglow; (2) formulae governing the initial electron distribution as it decouples from the shock and ceases to be accelerated; (3) photon production and absorption processes; and (4) a method of integrating flux from every location in the observable afterglow.  Several of these components have been presented in the literature previously \citep{GranotSari2002, ResslerLaskar2017} but we repeat them here for completeness.

The parameters needed for the model are those commonly used in the study of GRB afterglows.  The hydrodynamics of the jet are decided by the isotropic kinetic energy $E_\mathrm{iso}$ and the external density.  This density, $n_\mathrm{ext}(R) = A R^{-k}$, may depend on the distance from the central engine.  The choices $k = 0$ for a constant-density ambient medium and $k = 2$ for a wind-like ambient medium are typical.  Whichever value of $k$ is used, the parameter $A$ sets the number density of protons being encountered by the forward shock; for $k = 0$, $A \equiv n_\mathrm{ISM}$~cm$^{-3}$, while for $k = 2$, $A \equiv A_{\star}$~cm$^{-1}$.  Four parameters control the microphysics of the electron distribution and photon production.  The parameters $\epse$ and $\epsB$ are the fractions of inflowing kinetic energy placed in electrons and magnetic fields, respectively.\footnote{The remaining kinetic energy is placed in protons, such that $\epse + \epsB + \varepsilon_{p} = 1$.  Since protons are not significant contributors to photon production or absorption, we do not discuss them further in this work.}  The third microphysical parameter $p$ governs the steepness of the shock-accelerated electron distribution, i.e. $dN/dE \propto E^{-p}$. Finally, the fraction of electrons that are injected into the shock-acceleration process, and thus the fraction of electrons that form the non-thermal distribution, is controlled by the parameter $\fNT$.

Since these bursts occur at cosmological distances, our calculations use the redshift $z$ to the GRB.  In converting this to a luminosity distance we assume a flat Universe with $H_{0} = 67.7$~km~s$^{-1}$~Mpc$^{-1}$, $\Omega_{m} = 0.311$, and $\Omega_{\Lambda} = 1 - \Omega_{m}$ \citep{Wright2006, Planck2018VI}.

We make two further assumptions.  First of these is that the blast wave is assumed to be spherical: we do not consider the angular extent of the jet, nor any angular structure.  We cannot therefore discuss effects like a jet break, or the relationship between viewing angle and observations.  The second assumption is that we do not include a reverse shock in our model.  The physical state of the GRB ejecta, through which the reverse shock propagates, is far less certain than the physical state of the ambient medium.  As such the reverse shock is much less clearly defined than is the forward shock.  Including a reverse shock would mean the addition of new parameters beyond those listed in the previous paragraphs \citep{ZhangKobayashi2005,Japelj_etal_2014,BLL2021}.  Both the angular structure of the blast wave and any reverse shock are complications which we defer to future work.

\subsection{Hydrodynamics}
\label{sub:hydro}

Since we assume a spherically-symmetric burst, the bulk motion of the fluid is governed by the self-similar solution of \citet{BlandfordMcKee1976},
\begin{linenomath}\begin{equation}
  E_\mathrm{iso} = \frac{ 8\pi A m_{p} c^{2} \Gamma^{2} R^{3-k} }{ 17 - 4k } ,
  \label{eq:BM76_soln}
\end{equation}\end{linenomath}
for an adiabatic blast wave with a shock Lorentz factor $\Gamma$ at a distance $R$ from the central engine.  The radius of the shock is related to the time since the burst (in the rest frame of the progenitor) by
\begin{linenomath}\begin{equation}
  R = ct \left( 1 - \frac{ 1 }{ 2(4-k) \Gamma^{2} } \right) ,
  \label{eq:BM76_R_vs_t}
\end{equation}\end{linenomath}
and the radial location of points in the GRB interior may be expressed using the similarity variable
\begin{linenomath}\begin{equation}
  \chi = 1 + 2(4-k)\Gamma^{2}\left( \frac{ R - r }{ R } \right) .
  \label{eq:BM76_chi}
\end{equation}\end{linenomath}
In Equation~\ref{eq:BM76_chi}, and elsewhere in the paper, we use $R$ to refer to points on the forward shock (where $\chi = 1$) and $r$ to refer to points in the interior of the blast wave (where $\chi > 1$).  The similarity variable $\chi$ also relates the current state of the blast wave to the state at which a particular fluid parcel initially crossed the shock:
\begin{linenomath}\begin{equation}
  \chi = \left( \frac{ R }{ R_{0} } \right)^{4-k} ,
\end{equation}\end{linenomath}
where $R_{0}$ is the shock radius when $\chi = 1$ for the fluid under consideration.

The \citet{BlandfordMcKee1976} solution provides formulae for the proper energy density $e$, proper number density $n$, and the fluid Lorentz factor $\gamma$ in the progenitor rest frame:
\begin{linenomath}\begin{align}
  e &= 2 \Gamma^{2} A m_{p} c^{2} \chi^{-(17-4k)/3(4-k)} \nonumber \\
  n &= 2^{3/2} \Gamma A \chi^{-(10-3k)/2(4-k)} \nonumber \\
  \gamma &= 2^{-1/2} \Gamma \chi^{-1/2}
  \label{eq:BM76_hydro_vars}
\end{align}\end{linenomath}
We reserve $\Gamma$ for the Lorentz factor of the forward shock, and use $\gamma$ for the bulk motion of fluid.  As needed, the local magnetic field strength is computed using the equipartition parameter $\epsB$,
\begin{linenomath}\begin{equation}
  B^{2} = 8\pi \epsB e.
  \label{eq:mag_field}
\end{equation}\end{linenomath}
Note the assumption that $\epsB$ is uniform in space and constant in time.  It is expected from PIC simulations that there will be short-wavelength magnetic field turbulence at the forward shock, and that this turbulence can lead to detectable observational signatures \citep{Lemoine2013,LLW2013}.  At present we ignore this complication.

\subsection{Electron distribution}
\label{sub:elec_dist}

Rather than use PIC simulations \citep[e.g.,][]{SSA2013} or Monte-Carlo simulations \citep[e.g.,][]{WEBN2017} to determine the distributions of radiating particles, we define the distributions analytically based on the shock conditions when particles first encountered the blast wave, and on the history of the fluid post-shock.  We also track the evolution of the distributions as they advect away from the forward shock of the blast wave.

\subsubsection{The initial distribution}

As we are using a mixed distribution featuring both a thermal population and a non-thermal tail, we use four quantities to set the initial state of the distributions: (1) the location of the thermal peak; (2) the spectral index of the non-thermal tail; (3) the maximum energy, above which radiative losses exceed acceleration gains; and (4) the crossover point between the thermal peak and the non-thermal tail.  Since we consider only leptonic emission processes (see Section~\ref{sub:photons}), we need only discuss the electron distributions. 

The thermal component of the electron distribution function is given by the Maxwell--J{\"u}ttner distribution in momentum,
\begin{linenomath}\begin{equation}
  \frac{ d n_{e,\mathrm{TH}} }{ d\gamma_{e} } = C_\mathrm{TH} \gamma_{e}^{2} \beta_{e} e^{- \frac{ \gamma_{e} }{ a } } ,
  \label{eq:MJ_eqn}
\end{equation}\end{linenomath}
where $\gamma_{e}$ is the electron Lorentz factor in the plasma rest frame, and $\beta_{e} = \sqrt{1 - \gamma_{e}^{-2}}$. PIC simulations show that the thermal peak of the electron distribution lies at an energy $E = \epse \Gamma m_{p} c^{2} \gg m_{e} c^{2}$, from which we get the parameter $a$ in Equation~\ref{eq:MJ_eqn}:
\begin{linenomath}\begin{equation}
  a = \frac{ \epse \Gamma m_{p} }{ 3 m_{e} } ,
  \label{eq:MJ_a}
\end{equation}\end{linenomath}
the factor of 3 being necessary so that the average over the entire Maxwell--J{\"u}ttner distribution yields the desired value.

The non-thermal tail of the electron distributions is characterized by a fixed spectral index $p$ and a maximum Lorentz factor $\gamma_{e,\mathrm{max}}$:
\begin{linenomath}\begin{equation}
  \frac{ d n_{e,\mathrm{NT}} }{ d\gamma_{e} } = C_\mathrm{NT} \gamma_{e}^{-p} e^{ -( \gamma_{e} / \gamma_{e,\mathrm{max}} )^{2} } .
  \label{eq:NT_eqn}
\end{equation}\end{linenomath}
The maximum energy of the tail depends on the microphysics of shock acceleration and the strength of the radiative cooling.  We employ the empirical broken power law formula found in \cite{WBBN2021} for the maximum electron energy as a function of proper shock speed $\Gamma\beta$,
\begin{linenomath}\begin{equation}
  \gamma_{e,\mathrm{max}}(\Gamma\beta) = \gamma_{e,\mathrm{pk}} \left( \frac{ \Gamma\beta }{ \Gamma_\mathrm{pk}\beta_\mathrm{pk} } \right)^{L} \left[ \frac{ 1 + \frac{ L }{ H } }{ 1 + \left( \frac{ L }{ H } \right) \left( \frac{ \Gamma\beta }{ \Gamma_\mathrm{pk}\beta_\mathrm{pk} } \right) } \right]^{ \frac{ L + H }{ W } } .
  \label{eq:elec_max_en}
\end{equation}\end{linenomath}
This formula was derived from Monte Carlo simulations of electron acceleration and cooling in a microturbulent magnetic field, and so is well-suited for describing the electron distributions of GRB afterglows.  Equation~\ref{eq:elec_max_en} is controlled by five features,
\begin{linenomath}\begin{align}
  \gamma_{e,\mathrm{pk}} &= 6.35\xx{5} (m_{p}/m_{e}) \nonumber \\
  \Gamma_\mathrm{pk}\beta_\mathrm{pk} &= 0.532 (\epsB n_\mathrm{ext})^{-0.338} \nonumber \\
  L &= 2.12 \nonumber \\
  H &= 1.01 \nonumber \\
  W &= 5.26 (\epsB n_\mathrm{ext})^{0.11} .
  \label{eq:WBBN_features}
\end{align}\end{linenomath}
Briefly, the five features are: (1) $\gamma_{e,\mathrm{pk}}$, the Lorentz factor at the peak of the broken power law; (2) $\Gamma_\mathrm{pk}\beta_\mathrm{pk}$, the value of $\Gamma\beta$ at which the peak occurs; (3,4) $L$ (H), the power-law behavior of Equation~\ref{eq:elec_max_en} below (above) the peak; and (5) $W$, the width of the break.\footnote{The values presented in Equation~\ref{eq:WBBN_features} assume very efficient shock acceleration. Specifically, we have assumed Bohm-like diffusion, in which the electrons' mean free path through the magnetic field turbulence is equal to their gyroradius: $\lambda_\mathrm{mfp} = \eta_\mathrm{mfp} r_{g}$, with $\eta_\mathrm{mfp} = 1$.  If $\eta_\mathrm{mfp}$ were greater than 1, the maximum energy would be lower since electrons would generally spend more time completing each shock crossing cycle.} The physical meanings of these features, and their numerical values, are discussed at greater length in \citet{WBBN2021}.

Using the Blandford--McKee solution with the \citet{WBBN2021} maximum electron energy presents a problem.  The Blandford--McKee solution allows for arbitrarily large shock Lorentz factors, particularly at early times or for particles deep within the ``egg''.  Since the maximum electron energy is a decreasing function of Lorentz factor when $\Gamma \gg 1$, and the minimum electron energy is an increasing function of $\Gamma$, the model can predict electron distributions where the minimum energy is greater than the maximum energy.

\begin{figure}
  \epsscale{0.95}
  \includegraphics[width=0.5\columnwidth]{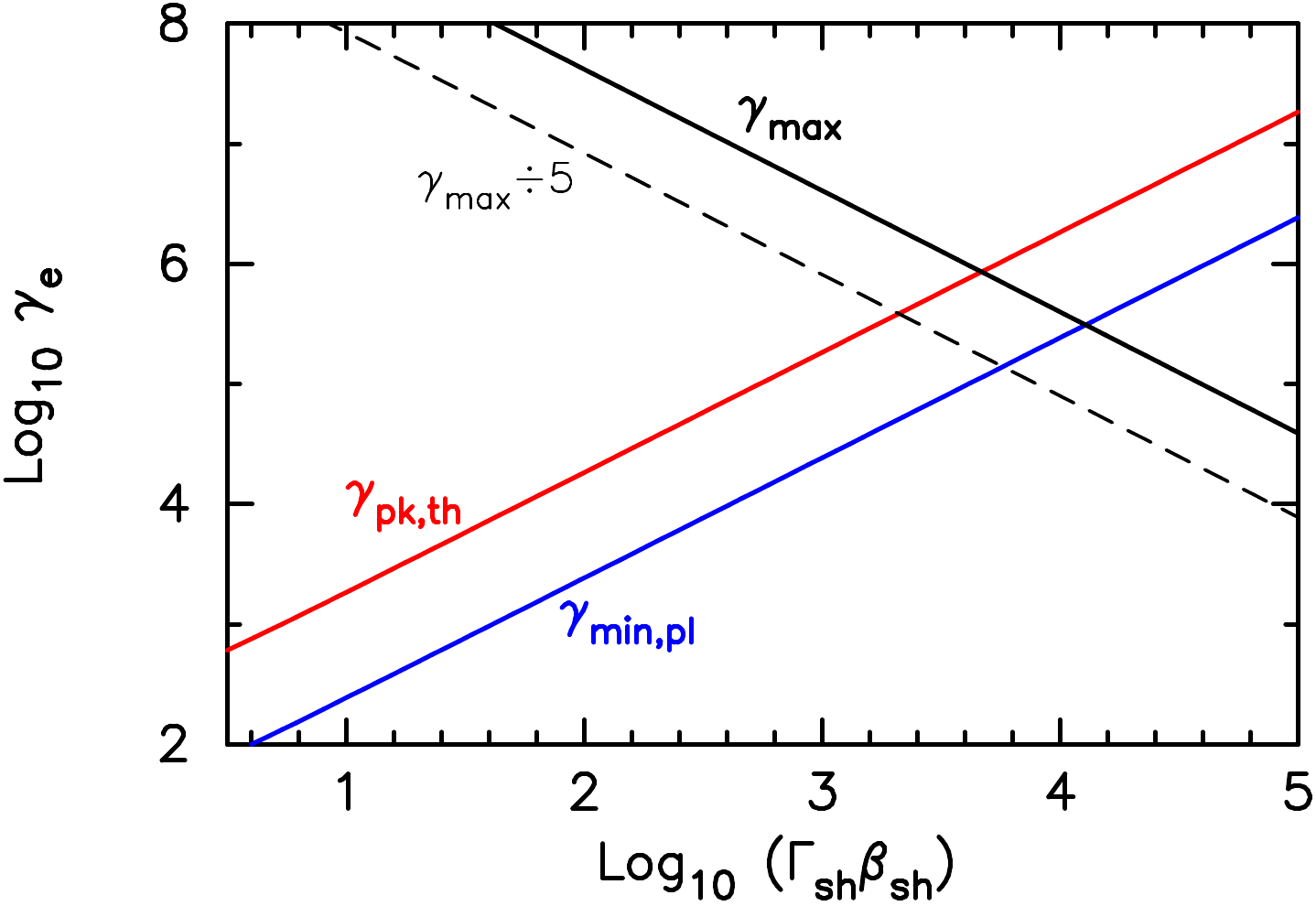}
  \caption{Key electron Lorentz factors as a function of the forward shock proper speed.  The red line is the location of the thermal distribution peak (Equation~\ref{eq:MJ_a}), and the blue line is the minimum electron Lorentz factor for a pure power law.  The solid black line is the maximum electron Lorentz factor predicted by Equation~\ref{eq:elec_max_en}.  The light, dashed black line below that is the same quantity, but shifted downward by a factor of 5.  All three electron Lorentz factors depend on the parameters of the GRB, which we list in Table~\ref{tab:GRB_params}.}
  \label{fig:elec_gam_minmaxes}
\end{figure}
The situation is illustrated in Figure~\ref{fig:elec_gam_minmaxes}.  The red and blue lines show possible ``minimum'' Lorentz factors for a nonthermal, shock-accelerated electron distribution: either the location of the thermal peak (in red), or the base of the power law when no thermal peak is present (in blue).  The thicker black line is the maximum electron Lorentz factor given by Equation~\ref{eq:elec_max_en}.  For sufficiently large shock speeds the maximum value of the shock-accelerated tail can be less than either plausible minimum value.  In order to cleanly separate the thermal peak from the nonthermal tail, one needs $\gamma_{e,\mathrm{max}} \gsim 5\gamma_{e,\mathrm{th,pk}}$ (see Figure~\ref{fig:elec_dists}).  This even more restrictive scenario is shown by the light dashed line in Figure~\ref{fig:elec_gam_minmaxes}.  The interpretation of such situations, where the maximum energy attainable by shock acceleration is less than (or close to) the energy of the thermal peak, is that no acceleration is possible (or that the accelerated population cannot be adequately distinguished from the thermal distribution).  If $\gamma_{e,\mathrm{max}} \ll \gamma_{e,\mathrm{th,pk}}$, one might even expect the inflowing electrons to significantly cool as they are heated by the microturbulent field, in which case the downstream electron distribution would no longer be thermal in nature.

In physical GRBs, however, the forward shock may not have arbitrarily large Lorentz factors.  It is generally held that a ``coasting'' phase takes place before deceleration begins, which would serve as an upper bound on the shock Lorentz factors used in our model \citep{MLR1993,KPS1999}.  We use a coasting Lorentz factor $\eta = 1000$ here, which is a plausible (if a bit high) upper limit and places $\gamma_{e,\mathrm{max}}$ well above the location of the thermal peak.

The final piece of the initial electron distribution is the crossover Lorentz factor $\gamma_{\times}$, where we stop using Equation~\ref{eq:MJ_eqn} and start using Equation~\ref{eq:NT_eqn}.  We calculate this in a manner similar to \citet{GianniosSpitkovsky2009} and \citep{ResslerLaskar2017} \citep[see also][]{KAB2019}: set the two equations to have an equal value at some electron energy, set their relative normalizations using the parameter $\fNT$, and set the total normalization of the combined distribution to the proper density (Equation~\ref{eq:BM76_hydro_vars}).  The three constraints are
\begin{linenomath}\begin{align}
  n ( 1 - \fNT ) &= \int_{1}^{\gamma_{\times}} C_\mathrm{TH} \gamma_{e}^{2}\beta_{e} e^{- \frac{ \gamma_{e} }{ a } } d\gamma_{e}, \nonumber \\
  n \fNT &= \int_{\gamma_{\times}}^{\infty} C_\mathrm{NT} \gamma_{e}^{-p} e^{ -( \gamma_{e} / \gamma_{e,_\mathrm{max}} )^{2} } d\gamma_{e}, \nonumber
\end{align}\end{linenomath}
and
\begin{linenomath}\begin{equation}
  C_\mathrm{NT} \gamma_{\times}^{-p} e^{ -( \gamma_{\times} / \gamma_{e,\mathrm{max}} )^{2} } = C_\mathrm{TH} \gamma_{\times}^{2}\beta_{\times} e^{- \frac{ \gamma_{\times} }{ a } } .
  \label{eq:norm_eqs}
\end{equation}\end{linenomath}
These three equations are solved numerically as needed, and provide not only the crossover energy $\gamma_{\times}$ but also the two normalization constants.  That is, given $n$, $\fNT$, $a$, and $\gamma_{e,\mathrm{max}}$ we compute $\gamma_{\times}$, which in turn allows us to find $C_\mathrm{TH}$ and $C_\mathrm{NT}$ and evaluate Equations~\ref{eq:MJ_eqn} and \ref{eq:NT_eqn}.

Note that if no thermal electrons are present, the electron distribution is defined by the pure power law of Equation~\ref{eq:NT_eqn}.  In such a situation the value of $C_\mathrm{NT}$ is fixed by the normalization condition,
\begin{linenomath}\begin{equation}
  C_\mathrm{NT} = n (p-1) \gamma_\mathrm{min}^{p-1}
  \label{eq:PPL_CNT}
\end{equation}\end{linenomath}
where
\begin{linenomath}\begin{equation}
  \gamma_\mathrm{min} = \frac{ (p-2) \epse \Gamma m_{p} }{ \sqrt{2} (p-1) m_{e} }
  \label{eq:PPL_gam_min}
\end{equation}\end{linenomath}
is the minimum electron Lorentz factor in the power law. It is worth pointing out that Equation~\ref{eq:PPL_gam_min} allows for $\gamma_\mathrm{min}$ values less than unity, if $p$ is approximately 2 and $\Gamma$ is not extremely large.  Care must be taken when applying this equation to ensure that one is not in regions of the parameter space where such an unphysical outcome can occur.  (The issue could be averted by casting Equations~\ref{eq:NT_eqn}, \ref{eq:PPL_CNT} and \ref{eq:PPL_gam_min} in terms of electron momentum rather than Lorentz factor.)

\subsubsection{Post-shock cooling}

The above equations define the electron distribution at the time electrons decouple from the shock---when the thermal electrons cross the shock, and when the shock-accelerated electrons cross the shock for the final time.  As the fluid advects downstream from the shock, the electrons experience both adiabatic and radiative cooling.  The plasma-frame Lorentz factors of the electrons evolve according to
\begin{linenomath}\begin{equation}
  \frac{ d \gamma_{e} }{ dt^{\prime} } = - \frac{ \sigma_{T} B^{2} \gamma_{e}^{2} }{ 6\pi m_{e} c } + \frac{ \gamma_{e} }{ 3 n }\frac{ dn }{ dt^{\prime} } ,
  \label{eq:elec_dgamdt}
\end{equation}\end{linenomath}
where the first term represents the radiative losses (with $\sigma_{T}$ being the Thomson cross section) and the second term represents the adiabatic losses.

Equation~\ref{eq:elec_dgamdt} can be rewritten in terms of the Blandford--McKee self-similarity variable $\chi$ \citep[for details, see][]{GranotSari2002}.  At a location $\chi$ within the shock structure, a hypothetical electron with infinite initial energy will have cooled to
\begin{linenomath}\begin{equation}
  \gamma_{\infty}(\chi) = \frac{ 2 (19 - 2k) \pi m_{e}^{2} c^{3} \gamma_{0} }{ \sigma_{T} B_{0}^{2} t_{0} } \frac{ \chi^{ (25-2k)/[6(4-k)] } }{ \chi^{ (19-2k)/[3(4-k)] } - 1 } .
  \label{eq:elec_max_cooled}
\end{equation}\end{linenomath}
In this equation $B_{0}$ and $t_{0}$ are, respectively, the magnetic field and progenitor-frame time at which the electrons began to cool, while $\gamma_{0}$ is the fluid bulk Lorentz factor (not the initial electron Lorentz factor; see Equation~\ref{eq:BM76_hydro_vars}) at time $t_{0}$.  Note that $\gamma_{\infty}(\chi)$ is simply an intermediate step in the calculation; since our electron distribution does not actually extend to infinite energy (being limited to approximately the value shown in Equation~\ref{eq:elec_max_en}) there are never electrons present with this Lorentz factor.  Electrons that did not possess infinite energy at $\chi = 1$ will, at some later time, have cooled to a Lorentz factor given by
\begin{linenomath}\begin{equation}
  \gamma_{e}(\gamma_{e,0}, \chi) = \frac{ \gamma_{e,0} }{ \chi^{ (13-2k)/[6(4-k)] } + \gamma_{e,0}/\gamma_{\infty}(\chi) } .
  \label{eq:elec_cooled}
\end{equation}\end{linenomath}

As pointed out in \citet{GranotSari2002}, the quantity $(dn_{e}/d\gamma_{e})(d\gamma_{e}/n)$ is conserved as the distributions evolve.  So to compute $dn_{e}/d\gamma_{e}$ at arbitrary $\chi$ we compute the change in $d\gamma_{e}$ using Equation~\ref{eq:elec_cooled}, then rescale by a further factor of $n/n_{0}$:
\begin{linenomath}\begin{equation}
  \frac{ dn_{e}(\chi) }{ d\gamma_{e} } = \frac{ dn_{e,0} }{ d\gamma_{e,0} } \frac{ d\gamma_{e,0} }{ d\gamma_{e} } \frac{ n }{ n_{0} } ,
  \label{eq:dndE_cooled}
\end{equation}\end{linenomath}
where the subscript $0$ denotes the initial values of the various quantities.

\subsection{Photon processes}
\label{sub:photons}

We consider three photon processes in this work, all leptonic in nature: synchrotron radiation, synchrotron self-absorption (SSA), and synchrotron self-Compton (SSC).  We do not consider hadronic emission processes (e.g. proton synchrotron or photopion production); these are many orders of magnitude less efficient than leptonic processes at a given photon energy, unless extreme values for the proton distribution or bulk fluid properties are assumed \citep{BottcherDermer1998, ZhangMeszaros2001, WEBL2015}.

Synchrotron radiation is computed using the traditional \citet{RybickiLightman1979} formula,
\begin{linenomath}\begin{equation}
  P_{\nu}(\nu, \gamma) = \frac{ \sqrt{3} q^{3} B \sin \alpha }{ m_{e} c^{2} } F\left( \frac{ \nu }{ \nu_{ch} } \right) 
\end{equation}\end{linenomath}
for an electron radiating in a magnetic field of strength $B$ at a pitch angle $\alpha$ between the magnetic field and the motion of the electron.  The characteristic frequency of such a photon is given by
\begin{linenomath}\begin{equation}
  \nu_{ch} = \frac{ 3 \gamma^{2} q B \sin \alpha }{ 4 \pi m_{e} c }
\end{equation}\end{linenomath}
where $\gamma$ is the Lorentz factor of the electron in the local rest frame of the plasma.  The function $F(x)$ is the usual synchrotron function,
\begin{linenomath}\begin{equation}
  F(x) = x \int_{x}^{\infty} K_{5/3} (t) dt ,
\end{equation}\end{linenomath}
in which $K_{5/3}$ is the modified Bessel function of the second kind, with order parameter $5/3$.  To find the total synchrotron power at a particular location, we average the emission over all pitch angles and integrate over the electron distribution:
\begin{linenomath}\begin{equation}
  P_{\nu,\mathrm{syn}} = \int \frac{ d n_{e}(\gamma) }{ d\gamma } P_{\nu,\mathrm{avg}}(\nu, \gamma) d\gamma
  \label{eq:P_syn}
\end{equation}\end{linenomath}

For low-energy photons, the synchrotron self-absorption (SSA) process can be a significant source of opacity.  The absorption coefficient is a function of both the electron distribution and the synchrotron power:
\begin{linenomath}\begin{equation}
  \alpha_{\nu} = -\frac{ 1 }{ 8 \pi m_{e} \nu^{2} } \int \gamma^{2} P_{\nu}(\nu, \gamma) \frac{ \partial }{ \partial\gamma }\left[ \frac{ d n_{e}/d\gamma }{ \gamma^{2} }\right] d\gamma
  \label{eq:SSA}
\end{equation}\end{linenomath}
When $h\nu \ll \gamma m_{e} c^{2}$, the function $P_{\nu}(\nu,\gamma)$ is proportional to $\nu^{1/3}\gamma^{-2/3}$, and the absorption coefficient can be computed analytically \citep{GPS1999ApJ527,Warren_etal_2018}.  In the interest of generality we do not make that assumption in this work, using Equation~\ref{eq:SSA} everywhere.

The third photon process we consider is SSC emission.  We use the procedure outlined in \citet{Jones1968}, which includes the Klein-Nishina reduction in scattering cross-section and is built upon the following equation:
\begin{linenomath}\begin{align}
  \frac{ d^{2}N }{ dt d\alpha_\mathrm{out} } &= \frac{ 2 \pi r_{0}^{2} c }{ \alpha_\mathrm{in} \gamma_{e}^{2} } \left[ 2 q^{\prime \prime} \ln q^{\prime \prime} + (1 + 2 q^{\prime \prime})(1 - q^{\prime \prime})  \right. \nonumber \\
  & \quad \quad \quad \quad \quad \quad \left. + \frac{ 1 - q^{\prime \prime} }{ 2 } \frac{ (4 \alpha_\mathrm{in} \gamma_{e} q^{\prime \prime})^{2} }{ 1 + 4 \alpha_\mathrm{in} \gamma_{e} q^{\prime \prime} } \right] ,
  \label{eq:FCJ_ssc}
\end{align}\end{linenomath}
where $dN/dt \approx c\sigma$ is the number of electron-photon collisions per unit time, normalized to the ambient photon density.  This equation calculates the number of photons produced per second per unit outgoing photon energy (expressed as a ratio of the electron rest mass, i.e. $\alpha_\mathrm{out} = E_{\gamma} / [m_{e} c^{2}]$).  The photons are assumed to be encountering a monoenergetic beam of electrons with Lorentz factor $\gamma_{e}$, and both $\alpha_\mathrm{out}$ and $\gamma_{e}$ are taken in the plasma rest frame.  The prefactor uses the classical electron radius, $r_{0} = q^{2}/[m_{e} c^{2}]$; the quantity $q^{\prime \prime} = \alpha_\mathrm{out} / [ 4 \alpha_\mathrm{in} \gamma_{e} (1 - \alpha_\mathrm{out}/\gamma_{e}) ]$ relates the incoming photon energy $\alpha_\mathrm{in}$ to the outgoing photon energy under consideration.

To convert Equation~\ref{eq:FCJ_ssc} into an SSC power, it is necessary to integrate over both the electron distribution and the number density of synchrotron photons present,
\begin{linenomath}\begin{equation}
  P_{\nu,\mathrm{ssc}} =   \frac{ h E_{\gamma,\mathrm{out}} }{ m_{e} c^{2} } \int_{1}^{\infty} d\gamma_{e} \frac{ dn_{e}(\gamma_{e}) }{ d\gamma_{e} } \int_{0}^{\alpha_\mathrm{out}} d\alpha_\mathrm{in} \frac{ dn_{\gamma}(\alpha_\mathrm{in} ) }{ d\alpha_\mathrm{in} } \frac{ d^{2} N }{ dt d\alpha_\mathrm{out} } ,
  \label{eq:P_ssc}
\end{equation}\end{linenomath}
where the prefactor in front of the integral converts $d^{2}N/(dt d\alpha_\mathrm{out})$ to $dP_{\nu}/d\nu_\mathrm{out}$.  Transforming the synchrotron $P_{\nu}$ given by Equation~\ref{eq:P_syn} into the photon density needed above requires an assumption about the size of the scattering region.  We assume that the region has a width equal to $ds$, the size of radiation transfer step as outlined in the next section.  With this assumption, we can relate $P_{\nu,\mathrm{syn}}$ to $d n_{\gamma}(\nu)/ d\nu$,
\begin{linenomath}\begin{equation}
  \frac{ d n_{\gamma}(\nu) }{ d\nu } = P_{\nu,\mathrm{syn}}~\frac{ ds }{ c \cdot h\nu } ,
  \label{eq:syn_phot_num_den}
\end{equation}\end{linenomath}
giving us everything necessary to compute $P_{\nu,\mathrm{ssc}}$.

\subsection{Radiative transfer}
\label{sub:rad_trans}

Radiative transfer is handled largely according to the method presented in \citet{GranotSari2002}.  That is, we solve the equation for radiative transfer in the rest frame of the GRB central engine:
\begin{linenomath}\begin{equation}
  \frac{ dI_{\nu}(\nu_\mathrm{GRB}) }{ ds } = j_{\nu,\mathrm{syn}}(\nu_\mathrm{GRB}) + j_{\nu,\mathrm{ssc}}(\nu_\mathrm{GRB}) - \alpha_{\nu}(\nu_\mathrm{GRB}) \, I_{\nu}(\nu_\mathrm{GRB}) .
  \label{eq:rad_trans_base}
\end{equation}\end{linenomath}
The quantities $j_{\nu,\mathrm{syn}}$ and $j_{\nu,\mathrm{ssc}}$ are related to Equations~\ref{eq:P_syn} and \ref{eq:P_ssc}, respectively, by $j_{\nu} = P_{\nu}/(4\pi)$, and must be computed in the plasma rest frame where emission occurs.  We use the Lorentz-invariant products $j_{\nu}/\nu^{2}$ and $\nu \alpha_{\nu}$ to express Equation~\ref{eq:rad_trans_base} in terms of plasma-frame quantities,
\begin{linenomath}\begin{equation}
  \frac{ dI_{\nu}(\nu_\mathrm{GRB}) }{ ds } = \mathcal{D}^{2} j_{\nu}(\nu_\mathrm{pf}) - \frac{ I_{\nu}(\nu_\mathrm{GRB}) \, \alpha_{\nu}(\nu_\mathrm{pf}) }{ \mathcal{D} } 
  \label{eq:rad_trans_used}
\end{equation}\end{linenomath}
where $\nu_\mathrm{pf} = \nu_\mathrm{Earth} (1+z) / \mathcal{D}$ is the plasma-frame photon frequency corresponding to $\nu_\mathrm{Earth}$ at Earth; the Doppler factor\footnote{As in Section~\ref{sub:hydro}, $\gamma$ and $\beta$ here refer to the fluid's bulk Lorentz factor and speed rather than to an individual particle's properties.} $\mathcal{D} = [\gamma ( 1 - \mu \beta )]^{-1}$ relates the plasma rest frame to the rest frame of the GRB progenitor; and the factor of $(1+z)$ transforms frequencies from the Earth frame to the GRB rest frame, such that $\nu_\mathrm{GRB} = \nu_\mathrm{Earth} (1+z)$.  Equation~\ref{eq:rad_trans_used} must be integrated along lines of sight through the entire volume emitting radiation observable at a time $\tobs$, a shape called the ``egg'' in \citet{GPS1999ApJ513}.  For a fuller discussion, see that paper, \citet{GPS1999ApJ527}, or especially \citet{GranotSari2002}.  Briefly, though, the shape is limited by the angle between a fluid parcel's position and the line of sight, as well as by the size and speed of the forward shock at any particular observer time:
\begin{linenomath}\begin{equation}
  R(\mu) = \frac{ c \tobs/(1+z) }{ 1 - \mu + [2(4-k)\Gamma(\mu)^{2}]^{-1} } .
  \label{eq:GPS_eggshell}
\end{equation}\end{linenomath}
When $R$ and $\Gamma$ are related by the Blandford--McKee solution, Equation~\ref{eq:GPS_eggshell} traces out an ovoid shape similar to its namesake.  The maximum perpendicular extent of the egg (that is, its angular size in the sky) varies as the shock decelerates, and is given by
\begin{linenomath}\begin{equation}
  R_{\perp,\mathrm{max}} = (5-k)^{(k-5)/[2(4-k)]} \frac{ R_\mathrm{axis} }{ \Gamma_\mathrm{axis} } ,
  \label{eq:GPS_egg_Rperpmax}
\end{equation}\end{linenomath}
where the subscripts on the right hand side denote quantities taken with $\mu = 1$.  To find the total emission at Earth, we integrate over the solid angle subtended by the egg.  Since we are assuming a spherical explosion and therefore azimuthal symmetry, it is convenient to recast the angular integral in terms of $x \equiv R_{\perp}/R_{\perp,\mathrm{max}}$, leading to
\begin{linenomath}\begin{equation}
  F_{\nu}(\nu_\mathrm{Earth}) = 2\pi (1+z) \left( \frac{ R_{\perp,\mathrm{max}}(\tobs) }{ d_{L} } \right)^{2} \int_{0}^{1} x \, I_{\nu}(\nu_\mathrm{GRB}) \, dx .
  \label{eq:Fnu_integral}
\end{equation}\end{linenomath}

Although we compute SSA in evaluating Equation~\ref{eq:Fnu_integral}, this is the only absorption process we consider.  SEDs presented in the remainder of this work ignore absorption due to the extragalactic background light (EBL), extinction due to gas and dust between the GRB and Earth, and other sources.

\section{Choosing $\fNT$}
\label{sec:fNT}

For the rest of this paper we discuss two realizations of the electron distribution.  The ``PPL'' model assumes that all electrons fall into a power-law distribution (Equations~\ref{eq:NT_eqn}, \ref{eq:PPL_CNT}, \ref{eq:PPL_gam_min}), and is the traditional assumption for the study of GRB afterglows.  The ``ThPL'' model uses the mixed thermal and nonthermal distribution described in Section~\ref{sub:elec_dist}.  Both the ThPL and PPL models use the same GRB parameters, and both distributions are normalized to the same local number density (which depends only on the hydrodynamic evolution and is not model-dependent).

The parameter $\fNT$ characterizes our ignorance of the actual microphysics present in and around turbulent relativistic shocks.  There is no theoretical lower limit on $\fNT$ (besides the obvious $\fNT \ge 0$), but there is a mathematical upper limit.  If $\gamma_{\times}$ falls below the thermal peak, then the combined distribution appears as a broken power law: rising until $\gamma_{\times}$, and falling afterward.  A simple computation \citep[presented in Appendix~\ref{sec:fNT_max} and confirmed by Equation~(17) in][]{ResslerLaskar2017} leads to $f_{\mathrm{NT},\mathrm{max}} = 3/(p+2)$.  When $p \approx 2.2-2.5$ as is traditionally assumed, $f_{\mathrm{NT},\mathrm{max}} \approx 0.7$; this value is far larger than any suggested by a physically-motivated model for particle acceleration, and is therefore merely a curiosity rather than a useful limitation on an uncertain parameter.

The so-called ``thermal leakage'' model \citep[e.g.,][]{Ellison1985,EWB2013,Warren_etal_2018} predicts particle injection at rates $\gsim 10$\%.\footnote{The plasma downstream from a strong relativistic shock recedes at speed of $c/3$; any particle whose parallel velocity component (i.e. along the shock normal) exceeds this value can counter its advection with the plasma and re-cross the shock if it is traveling upstream.  For a totally isotropic distribution of particles with speed $\approx c$, roughly 30\% of the particles satisfy this condition, and thus can enter the Fermi acceleration process.}  On the other hand, PIC simulations of relativistic shock formation and particle injection predict much smaller injection rates, $\sim 1-3$\% \citep{SSA2013, SKL2015}, for shocks propagating into an unmagnetized medium.  Shocks encountering a magnetized medium inject fewer particles still, down to and including $\fNT \approx 0$ for sufficiently strong upstream magnetic fields in the proper orientation \citep{SironiSpitkovsky2011, SSA2013}.

Despite the wide range of $\fNT$ allowed by PIC simulations, and the even larger range if one includes the thermal leakage model for electron injection, there is a natural choice if one wishes to determine the effects of a thermal population of electrons.  That choice, as applied previously in \citet{ResslerLaskar2017}, is to align the nonthermal portion of the ThPL distribution with the PPL distribution.  This is not a physically-motivated choice: ideally one would choose $\fNT$ based on the results of PIC simulations or analytic calculations specific to the shock speed and magnetic field structure.  The assumption we make here is entirely ad hoc; it is chosen to maximize the similarity between ThPL and PPL distributions rather than being based on more fundamental calculations.

\begin{figure}
  \epsscale{0.95}
  \includegraphics[width=0.5\columnwidth]{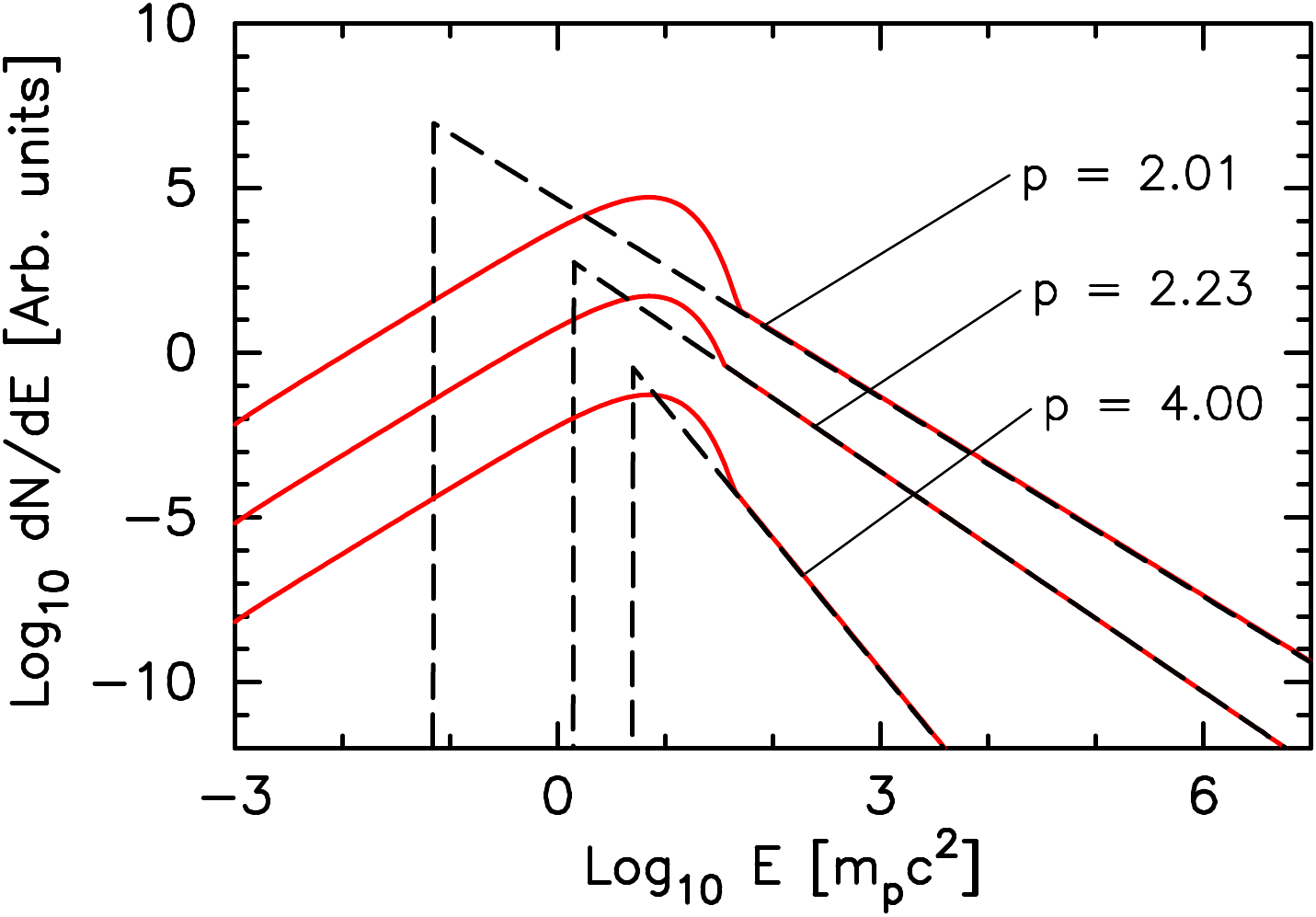}
  \caption{Electron distributions $dN/dE$ as functions of electron energy.  Dashed black lines show a pure power-law distribution with the specified index $p$.  Solid red lines show mixed thermal and nonthermal spectra, with $\fNT$ set so the nonthermal parts of the spectra overlap.  All spectra were calculated for a shock with $\Gamma_{0} = 100$ and $\epse = 0.1$, and a normalization of 1~cm$^{-3}$; distributions were then scaled to enhance readability.}
  \label{fig:elec_dists}
\end{figure}
We present in Figure~\ref{fig:elec_dists} several examples of electron distributions.  All distributions were computed using the same values of $\Gamma_{0}$, $\epse$, and total normalization.  The sole free parameter was the fraction $\fNT$ of the distribution in the nonthermal tail, which was set so that the PPL distribution overlapped the nonthermal portion of the ThPL distribution.  It is evident from the figure that the two kinds of electron distributions have maxima at different locations, and that this difference depends on $p$.  It can also be seen in the figure that the ``critical'' value of $\fNT$ required for overlap depends on $p$, as nonthermal particles make up a larger fraction of the overall distribution when $p = 2.23$ than they do in either of the other two cases ($p = 2.01$ and $p = 4$).

One can solve for $\fNTcrit$ by applying a fourth constraint to the system listed in Equation~\ref{eq:norm_eqs}.  The value of the power-law portion of the ThPL distribution must match the value of the PPL distribution at $\gamma_{\times}$, which sets $C_\mathrm{NT}$ to the value given in Equations~\ref{eq:PPL_CNT} and \ref{eq:PPL_gam_min}.  Since $\gamma_{\times} \ll \gamma_{e,\mathrm{max}}$, the exponential rollover in the power laws may be neglected, simplifying the equations considerably. The four equations ultimately reduce to the following transcendental equation,

\begin{linenomath}\begin{equation}
  2 e^{\delta} - 2 - 2 \delta - \delta^{2} = \frac{ 1 - \fNTcrit }{ (p-1) \fNTcrit} \delta^{3}
  \label{eq:fNT_crit}
\end{equation}\end{linenomath}
\citep[compare Equation~(17) in][]{ResslerLaskar2017} with $\delta$ defined as follows:
\begin{linenomath}\begin{equation}
  \delta(\fNTcrit) \equiv \frac{ \gamma_{\times} }{ a } = \frac{ 3 }{ \sqrt{2} } \frac{ (p-2) }{ (p-1) } \fNTcrit^{1/(1-p)}
\end{equation}\end{linenomath}

\begin{figure}
  \epsscale{0.95}
  \includegraphics[width=0.5\columnwidth]{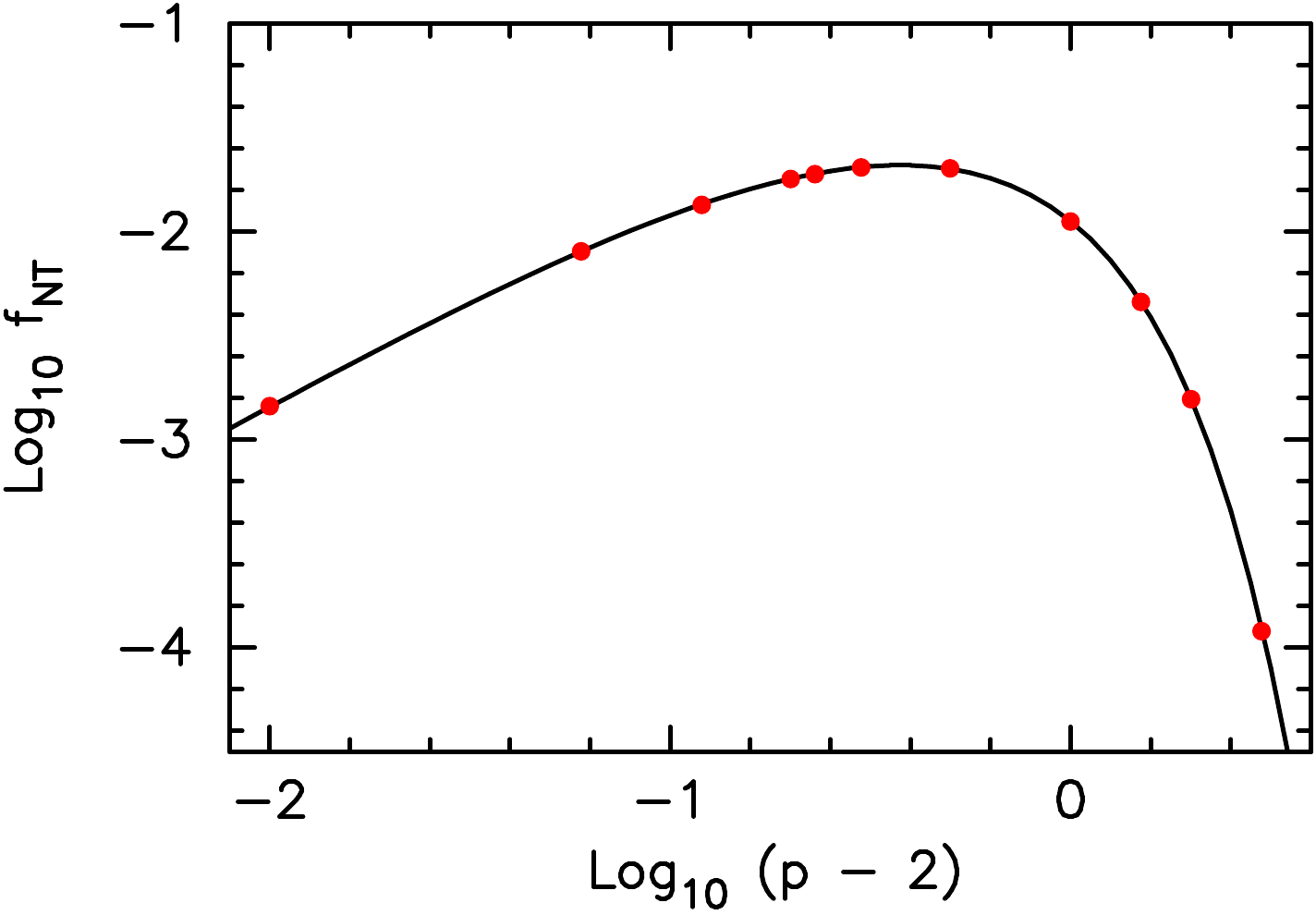}
  \caption{Value of $\fNT$ required for overlap in Figure~\ref{fig:elec_dists}, as a function of nonthermal index $p$.  The red dots are taken from Table~\ref{tab:fnt_vs_p}, while the black curve is the solutions to Equation~\ref{eq:fNT_crit}.}
  \label{fig:p_fnt-vs-p}
\end{figure}

Figure~\ref{fig:p_fnt-vs-p} shows the solutions to Equation~\ref{eq:fNT_crit} over a range of $p$ with astrophysical relevance.  Selected points are listed for reference in Table~\ref{tab:fnt_vs_p}.  Both the figure and the table show a trend where $\fNTcrit$ is small for $p \approx 2$ and $p \gsim 4$, with an apparent peak around $p \approx 2.3$.  When $p \rightarrow 2$ the base of the power law, $E_\mathrm{min}$, goes to zero.  When $p$ is large, the spectrum drops off so steeply that the portion lying between $E_\mathrm{min}$ and $E_{\times}$ (refer back to Equation~\ref{eq:norm_eqs}) is an increasing fraction of the power law's normalization.  In both limits the portion of the power law above $E_{\times}$ is expected to decrease, and this expectation is borne out by the data.  We note also that $\fNTcrit$ is independent of the shock Lorentz factor $\Gamma$ and the electron equipartition factor $\epse$.  Since the values of both $E_\mathrm{min}$ and the thermal peak depend in the same way on $\Gamma$ and $\epse$, changing either parameter merely shifts the curves in Figure~\ref{fig:elec_dists} left or right; it does not affect the location of the intersection point relative to either distribution's maximum, and therefore the value of $\fNTcrit$ is also unaffected.

\begin{table}[h]
\centering
\begin{tabular}{ll}
\hline
$p$ & $\fNTcrit$ \\
\hline
2.01 & $1.44\xx{-3}$ \\
2.06 & $8.06\xx{-3}$ \\
2.12 & $1.37\xx{-2}$ \\
2.20 & $1.76\xx{-2}$ \\
2.23 & $1.90\xx{-2}$ \\
2.30 & $2.12\xx{-2}$ \\
2.50 & $1.97\xx{-2}$ \\
3.00 & $1.06\xx{-2}$ \\
3.50 & $4.87\xx{-3}$ \\
4.00 & $1.54\xx{-3}$ \\
5.00 & $1.12\xx{-4}$ \\
\hline
\end{tabular}
\caption{\label{tab:fnt_vs_p} Normalization constant $\fNTcrit$ associated with each spectral index $p$.}
\end{table}

Given the many orders of magnitude spanned by $\fNT$ in PIC simulations, it is rather unexpected that the predictions for unmagnetized shocks ($\sim 1-3\%$) align so closely with the values listed in Table~\ref{tab:fnt_vs_p} for typical shock-accelerated distributions with $p \approx 2.2$.  The coincidence is all the more striking because Table~\ref{tab:fnt_vs_p} relied on the unphysical assumption of simply matching non-thermal populations with and without a thermal peak present.

\section{SSC and SSA with the full distribution}
\label{sec:full_dist}

Here, and for the rest of the paper, we choose a fiducial set of GRB parameters to compute the evolution of the afterglow.  These values are $E_\mathrm{iso} = 10^{52}$~erg, $k = 0$, $A = n_\mathrm{ism} = 1$~cm$^{-3}$, $\epse = 0.1$, $\epsB = 0.01$, and $p = 2.23$.  When discussing the ThPL model, the parameter $\fNT$ is set equal to 0.019, for reasons we elaborated on in Section~\ref{sec:fNT}.  The maximum Lorentz factor attained by the GRB forward shock is $\eta = 1000$.  The burst is assumed to take place at a redshift $z = 1$.  The full set of fiducial parameters is listed in Table~\ref{tab:GRB_params}.  We require that $\Gamma > 2.5$ everywhere since the Blandford--McKee solution ceases to be an adequate description of the hydrodynamics at lower shock Lorentz factors \citep{KPS1999}; for the above parameters, the latest observer time we can reasonably discuss is $\tobs \approx 1.2\xx{6}$~s.\footnote{The short-wavelength magnetic field turbulence in GRB afterglows is seeded by plasma instabilities that quench at $\Gamma_{0} \sim 10$ \citep{LemoinePelletier2011}.  It is possible that the magnetic field properties of the blast wave (and the associated values of $\epse$ and $\epsB$) will change long before the fluid solution ceases to apply.  For now, we ignore this likely complication in the interest of simplifying the model.}
\begin{table}[h]
\centering
\begin{tabular}{ll}
\hline
Parameter & Value \\
\hline
$E_\mathrm{iso}$ & $10^{52}$~erg \\
$k$ & $0$ \\
$A$ $(=n_\mathrm{ism})$ & $1$~cm$^{-3}$ \\
$\epse$ & $0.1$ \\
$\epsB$ & $0.01$ \\
$p$ & $2.23$ \\
$\fNT\phantom{.}^{a}$ & $1.9\xx{-2}$ \\
$\eta$ & $1000$ \\
$z$ & $1$ \\
\hline
$H_{0}$ & $67.7$~km~s$^{-1}$~Mpc$^{-1}$ \\
$\Omega_{m}$ & $0.311$ \\
$\Omega_{\Lambda} (=1-\Omega_{m})$ & $0.689$ \\
\hline
\end{tabular}
\caption{\label{tab:GRB_params} List of input GRB parameters and their fiducial values.}
\footnotesize{$^{a}$ See discussion in Section~\ref{sec:fNT}.}
\end{table}

We now include the full suite of processes discussed in Section~\ref{sec:model}.  In this section we discuss the impacts of SSA and SSC, as well as limiting the maximum electron energy to physically-plausible values.  We remind the reader that none of the SEDs presented here include intergalactic absorption due to the EBL.

\begin{figure}
  \epsscale{0.95}
  \includegraphics[width=0.5\columnwidth]{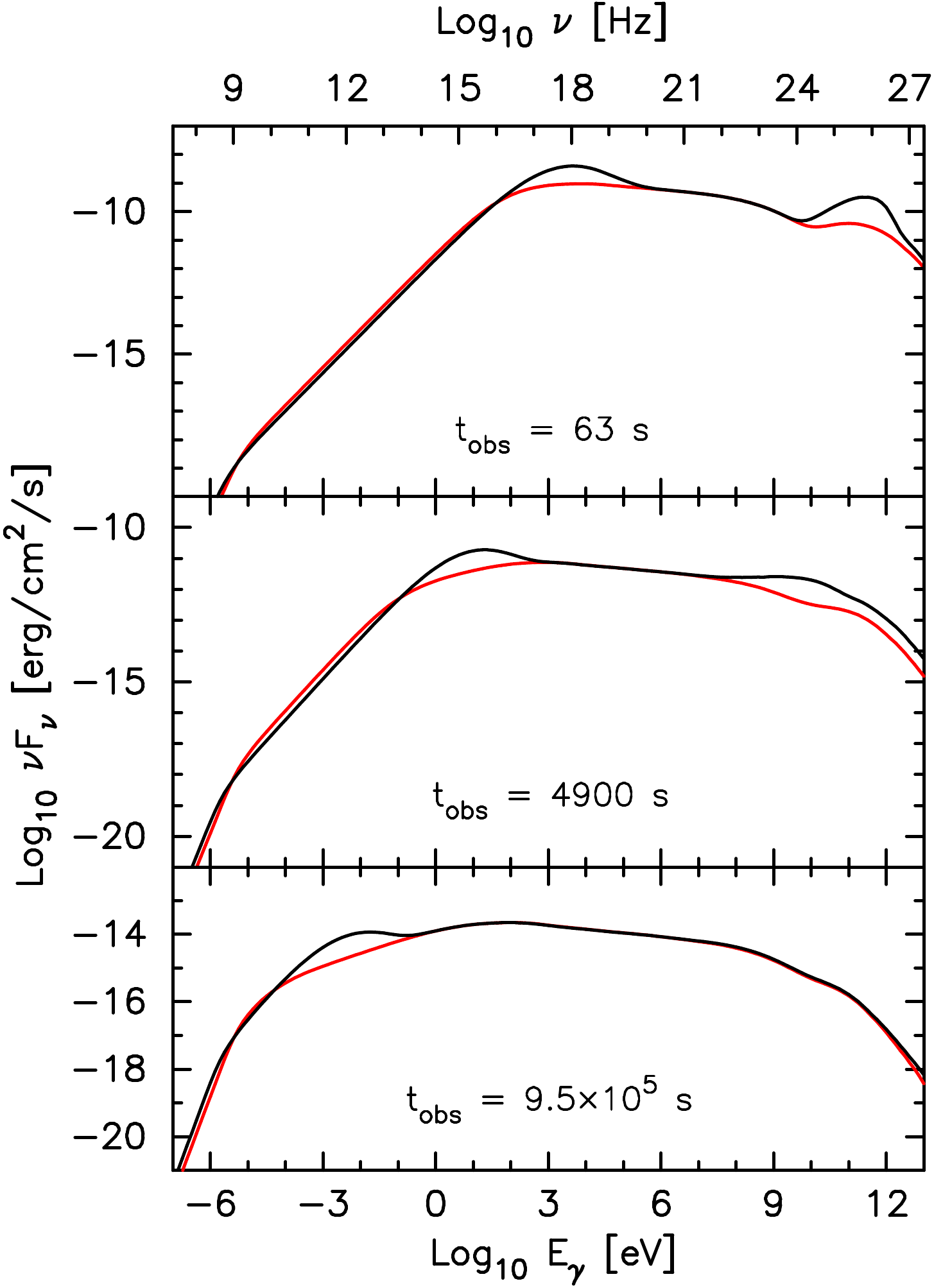}
  \caption{Comparison of SEDs at three observer times, at photon energies from 24~MHz to upwards of 1~TeV.  The black line is that of the ThPL model, while the red line is from the PPL model.  Note the differing vertical axes for the three subpanels.}
  \label{fig:p_pl-therm_ssc_compare}
\end{figure}

\begin{figure}
  \epsscale{0.95}
  \includegraphics[width=0.5\columnwidth]{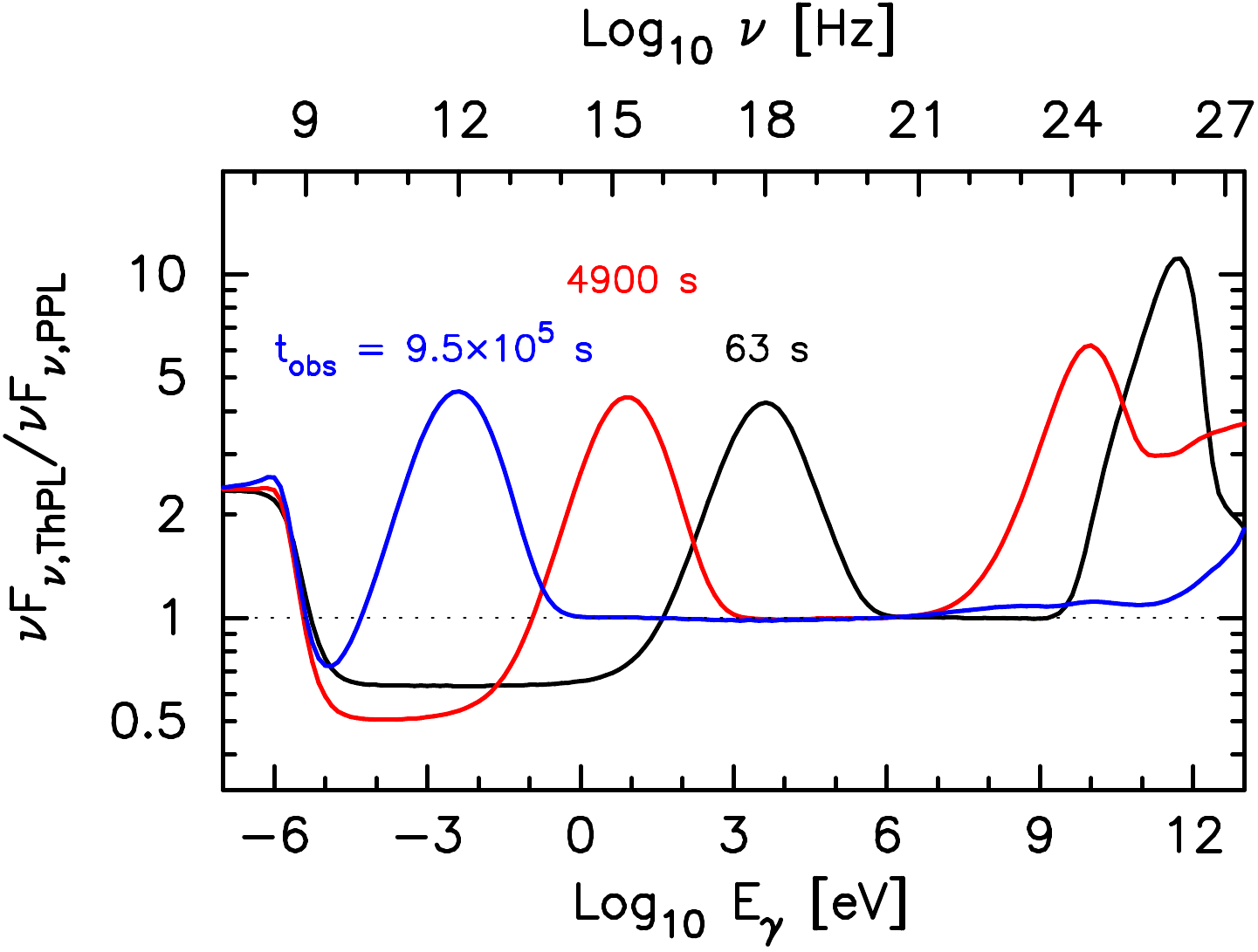}
  \caption{Ratio of the ThPL flux to the PPL flux at the three observer times in Figure~\ref{fig:p_pl-therm_ssc_compare}.  The ratios for the three times are plotted in different colors, with black for the SEDs at $\tobs = 63$~s, red for $\tobs = 4900$~s, and blue for $\tobs = 9.5\xx{5}$~s.}
  \label{fig:p_SEDs_ratio}
\end{figure}

We first compare SEDs from the ThPL (with thermal particles) and PPL (without thermal particles) models in Figure~\ref{fig:p_pl-therm_ssc_compare}, to illustrate the impact of thermal electrons at virtually every part of the observable spectrum.  The ratios of SED pairs are presented in Figure~\ref{fig:p_SEDs_ratio}, offering further insight into the differences between the two distributions.

There is a substantial difference between the ThPL and PPL distributions (see Figure~\ref{fig:elec_dists}), and this difference unsurprisingly carries over to the photon spectra.
\begin{itemize}
  \item In the radio, there is some difference between the two models, with the ThPL model showing slightly enhanced emission below $\sim$GHz frequencies throughout the afterglow.  The enhancement would be even greater if we were comparing the ThPL model against the fitting formulae of \citet{GranotSari2002}, due to our correct treatment of SSA and the lower value of $\nu_{a}$ here versus there (see Figures~\ref{fig:p_GS2002_code_check} and \ref{fig:p_GS2002_code_check_hi-res_betas_zoom} in Appendix~\ref{sec:alpha_nu}).
  \item At photon energies between infrared and X-ray, either of the ThPL or PPL distributions could result in more photon flux.  Looking back at Figure~\ref{fig:elec_dists}, PPL electrons outnumber ThPL electrons at the base of the pure power law.  At slightly higher electron energies the thermal peak of the ThPL model contains more electrons.  This inversion also shows up in the SEDs: as long as the characteristic synchrotron energy of the thermal peak is above any particular waveband, the base of the PPL distribution produces more flux than does the $\gamma_{e}^{2}$ tail of the thermal electron population.  Eventually the thermal peak shifts to low enough energy that its synchrotron flux dominates that of the power law, leading to the obvious peaks in the SEDs at each observer time in Figures~\ref{fig:p_pl-therm_ssc_compare} and \ref{fig:p_SEDs_ratio}.
  \item Recall that $\fNT$ was chosen specifically so that the accelerated, nonthermal portion of the ThPL distribution matched the PPL distribution (Section~\ref{sec:fNT}). It is to be expected that the SEDs also overlap above the synchrotron thermal peak, which is indeed the behavior the two models show.  For most of the X-ray afterglow, then, there is not much difference between the two electron distributions.
  \item At GeV and TeV energies there is again a very clear difference between the ThPL and PPL SEDs, due here to the ``thermal-thermal'' SSC peak: synchrotron photons both produced by, and upscattered by, thermal electrons.  The difference between the two models is upwards of an order of magnitude early in the afterglow, but it fades with time until there is only marginal enhancement due to thermal particles.  Above $\sim 1$~TeV the ThPL model is persistently enhanced compared to the PPL model, but photon fluxes at these energies are already so low---even without EBL absorption considered---that there is little observational relevance to the distinction.
\end{itemize}
The characterization of the two models at X-ray energies and below agrees qualitatively with prior results, in particular those of \citet{ResslerLaskar2017}.  Differences between our work and theirs are likely due to how we treated the thermal distribution.  While we fixed the energy of our thermal peak and then found $\gamma_{\times}$ based on a chosen $\fNT$, \citet{ResslerLaskar2017} computed $\gamma_{\times}$ (called $\gamma_{\mathrm{min},0}$ there) first and then fixed the thermal peak using $\fNT$ and $\gamma_{\times}$.
\begin{figure}
  \epsscale{0.95}
  \includegraphics[width=0.5\columnwidth]{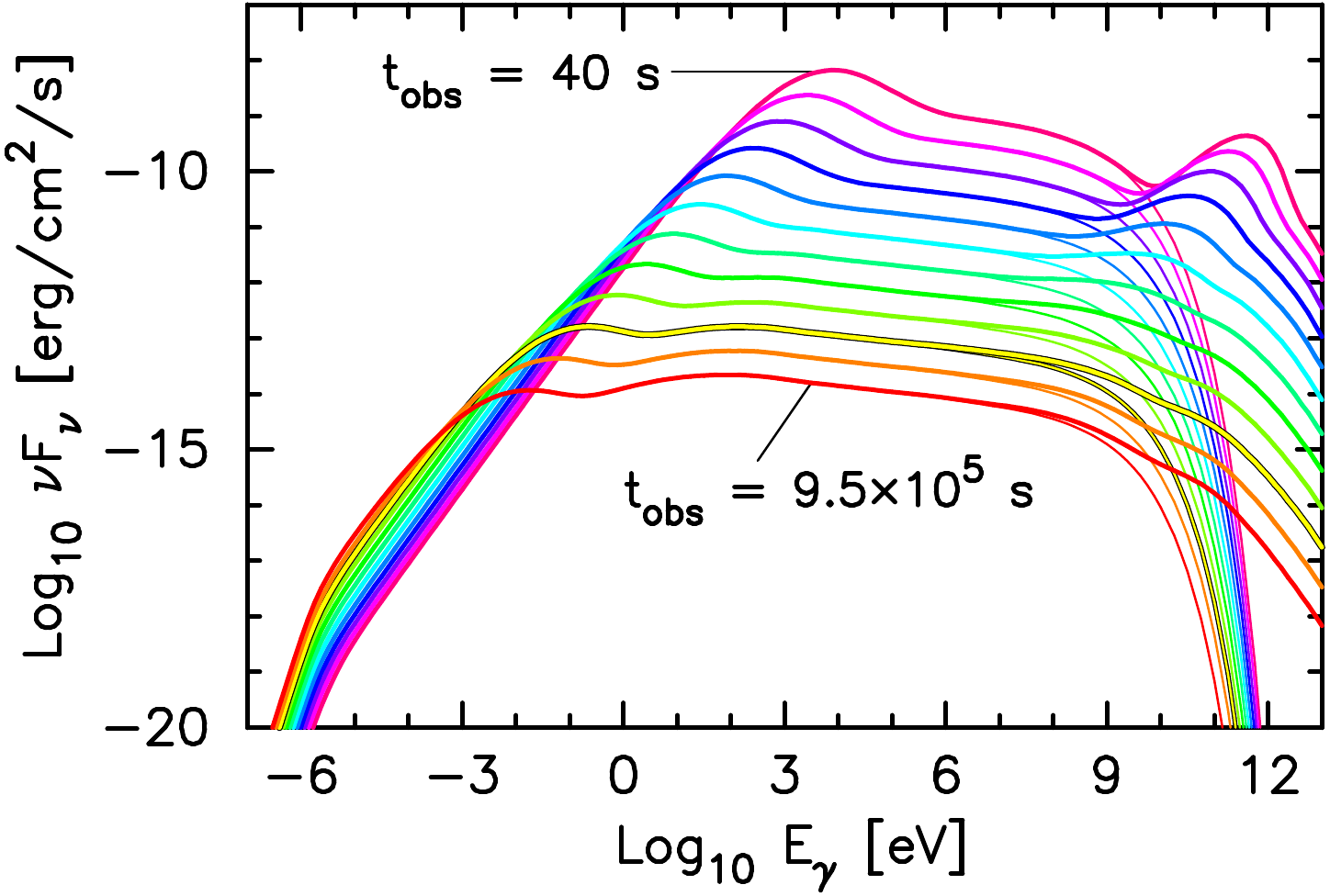}
  \caption{Full SEDs for various times throughout the hypothetical afterglow.  The earliest time, in deep pink, is 40~s.  Successive SEDs are separated by a factor of 2.5 in observer time, until the red line at $\tobs = 9.5\xx{5}$~s.  At each time, the thick line traces out the full SED, while the thin line is the contribution of synchrotron radiation.}
  \label{fig:p_SEDs_time_sequence}
\end{figure}
We now shift our focus away from model comparison and towards a full simulated afterglow using the ThPL model and the parameters listed in Table~\ref{tab:GRB_params}.  In Figure~\ref{fig:p_SEDs_time_sequence} we plot the SEDs for a sequence of times, from $\tobs = 40$~seconds to $\tobs \approx 11$~days.  The figure also shows SEDs due solely to synchrotron emission, i.e. without SSC.  The synchrotron-only curves begin the exponential decay associated with $\gamma_{e,\mathrm{max}}$ (Equation~\ref{eq:NT_eqn}) at about 1~GeV at every observer time computed, which is in excellent agreement with Equation~(24) of \citet{WBBN2021}.  At the opposite end of the SED, the $F_{\nu} \propto \nu^{1/3}$ segment is visible at all times, confirming that the afterglow is in the slow-cooling regime from the earliest observer time shown.  The transition from fast to slow cooling happens at
\begin{linenomath}\begin{align}
  t_\mathrm{trans} &\sim 10^{2}~\mathrm{sec}~(1+z) \frac{ (p-2)^{2} }{ (p-1)^{2} } \varepsilon_{e,-1}^{2} \varepsilon_{B,-2}^{2} n_\mathrm{ext,0} E_{52} \nonumber \\
  &\sim 7~\mathrm{sec}
  \label{eq:t_fastslow}
\end{align}\end{linenomath}
for our fiducial parameters \citep{GranotSari2002}.  This is before any observer time presented in Figure~\ref{fig:p_SEDs_time_sequence}; indeed, it is earlier than any observer time considered in this work.

\begin{figure}
  \epsscale{0.95}
  \includegraphics[width=0.5\columnwidth]{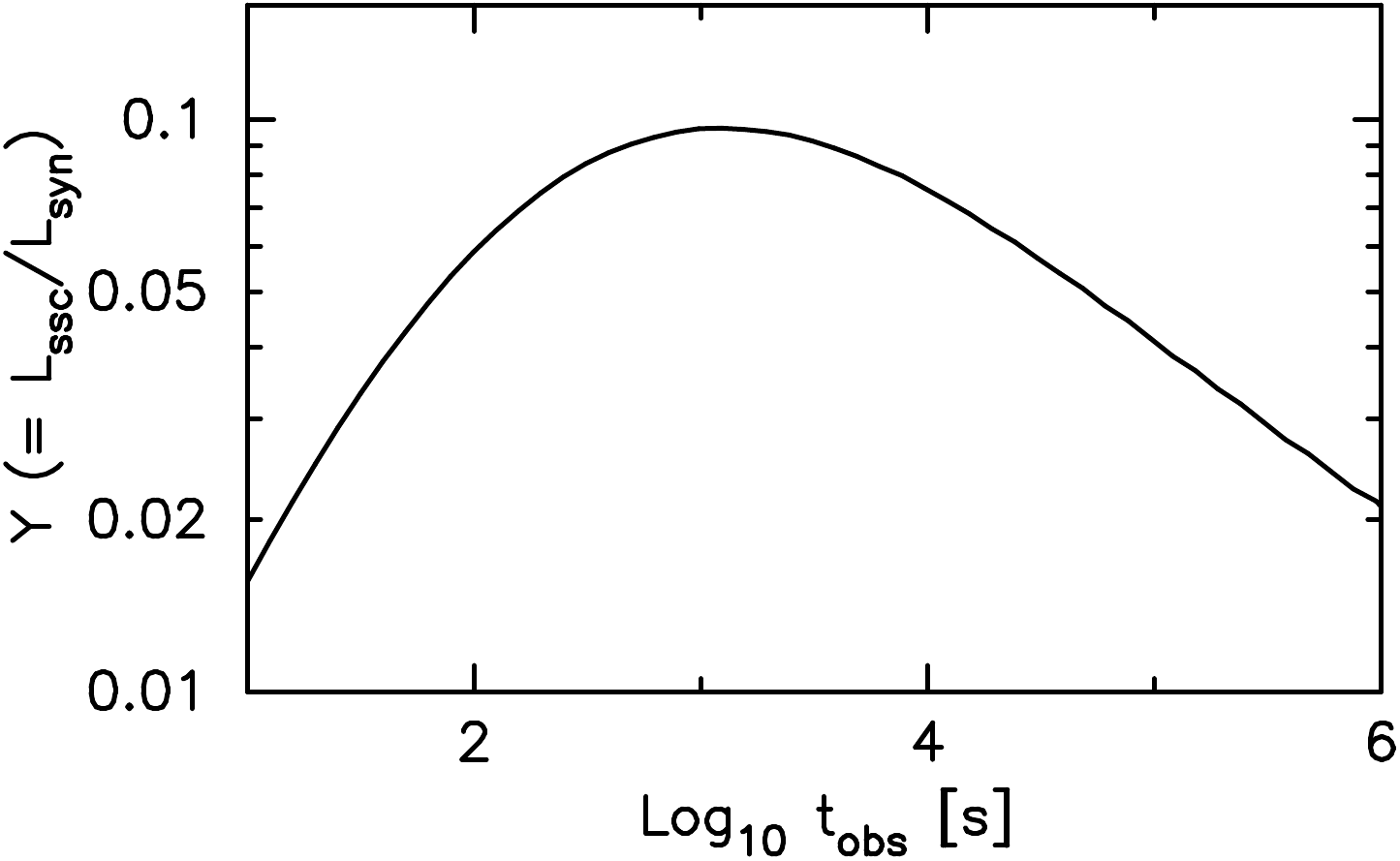}
  \caption{Evolution of the global Compton parameter as a function of observer time.}
  \label{fig:p_compton_Y}
\end{figure}
The process outlined in Section~\ref{sec:model} allows us to compute the Compton $Y$ parameter \citep[$= L_{\mathrm{ssc}}/L_{\mathrm{syn}}$,][]{SariEsin2001} directly.  Not only do we determine $j_{\nu,\mathrm{syn}}$ and $j_{\nu,\mathrm{ssc}}$ locally everywhere in the emitting volume, but we can use the SEDs in Figure~\ref{fig:p_SEDs_time_sequence} to discuss the global values of $\int F_{\nu,\mathrm{syn}} d\nu$ and $\int F_{\nu,\mathrm{ssc}} d\nu$ and how their ratio evolves over the course of the simulated afterglow.  This ratio is hinted at in Figure~\ref{fig:p_SEDs_time_sequence}, but explicitly presented in Figure~\ref{fig:p_compton_Y}.  At no point in our simulated afterglow does $Y$ exceed, or even approach, unity: the luminosity due to SSC is a small fraction of the synchrotron luminosity over the entire time considered.   The curve in Figure~\ref{fig:p_compton_Y} is the global value across the entire shocked region, and it is possible that $Y$ can vary locally.  With that caveat, though, to a first approximation it is reasonable to neglect the backreaction of SSC on electron cooling \citep[compare against][for the scenario where $Y$ is too large to be neglected]{ZachariasSchlickeiser2013}.  This retroactively justifies the assumption made in \citet{WBBN2021} to ignore SSC as a cooling process, for at least some part of the parameter space relevant to GRB afterglows. 

Since our treatment of SSC in GRB afterglows is a novel extension to the existing literature, we wish to expand and reiterate on a few points.  The effects of the thermal population on SSC are most important early in the afterglow and at energies above $\sim$GeV.  In this time and energy range, the difference between the ThPL and PPL models can be more than an order of magnitude even for the parameters listed in Table~\ref{tab:GRB_params}---which were not chosen for the purpose of maximizing the SSC peaks in Figures~\ref{fig:p_pl-therm_ssc_compare} and \ref{fig:p_SEDs_ratio}.  Though we defer such treatment until future work, one can imagine that there are regions of the parameter space where the difference between the ThPL and PPL models is even greater in magnitude, or lasts even later into the afterglow.

The synchrotron contribution to the SEDs ends at roughly 1~GeV for all observer times.  Given the steepness of the exponential rolloff, virtually all photons detected above 1~GeV are therefore produced by the SSC mechanism.  This is in line with previous predictions for GeV and TeV emission from GRB afterglows \citep{MRP1994, ZhangMeszaros2001, LWW2013, Wang_etal_2019, Fraija_etal_2019ApJ883, MAGIC2019Natur575, Abdalla_etal_2019Natur575,DerishevPiran2016,DerishevPiran2019}.  Emission at very high energies is further enhanced when a thermal distribution of electrons is present, resulting in TeV afterglows that are (1) brighter at a given observer time, and/or (2) detectable for longer into the afterglow.

\section{Light curves, spectral and temporal indices}
\label{sec:LCs_indices}

In the previous section we discussed the behavior of multi-wavelength SEDs at individual observer times.  Now we consider the behavior over individual energy ranges and how they evolve with time.  Throughout this, and following, sections, we adopt the convention $F_{\nu} \propto t^{-\alpha} \nu^{-\beta}$. 

\begin{figure}
  \epsscale{0.95}
  \includegraphics[width=0.5\columnwidth]{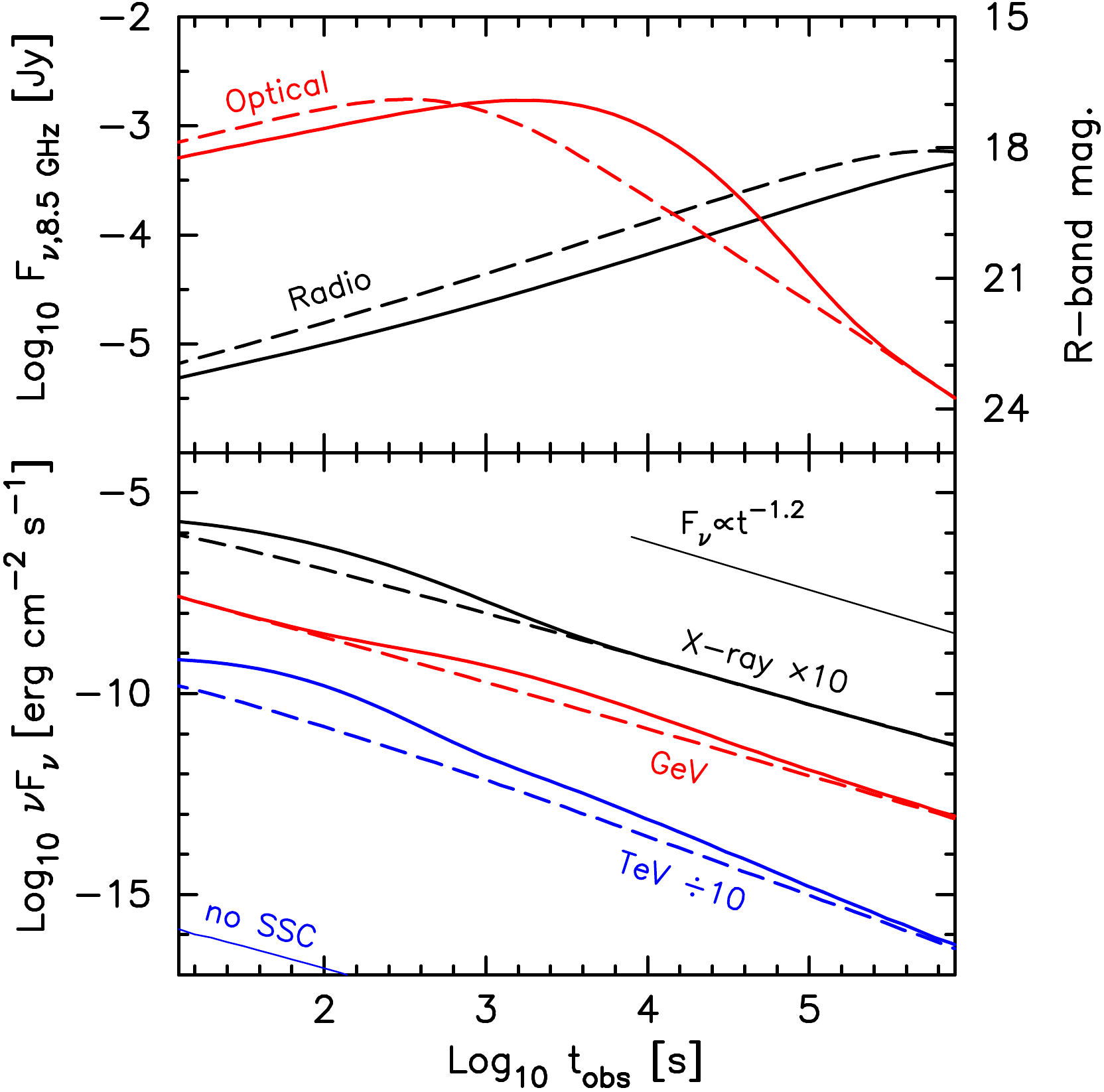}
  \caption{Light curves at five energies.  In both panels, solid lines are light curves produced by the ThPL model; dashed lines were computed using the PPL model.  \textit{Top panel}: light curves in radio (8.46 GHz; left axis) and optical (640 nm; $\sim$R-band; right axis).  \textit{Bottom panel}: light curves in X-ray ($0.3-10$ keV, \textit{Swift} XRT), GeV ($80-25000$ MeV, \textit{Fermi} LAT), and TeV ($300-1000$ GeV, MAGIC).  The X-ray and TeV curves have been scaled for clarity reasons, and the thin line at the bottom left is TeV emission (scaled like the other TeV curve) if SSC is ignored as a photon production process.}
  \label{fig:p_phot_LCs}
\end{figure}
The light curves at five energies of interest are shown in Figure~\ref{fig:p_phot_LCs}, and we compare the behavior of afterglows based on the ThPL and PPL models.
\begin{itemize}
  \item At radio wavelengths above $\nu_{a}$ the ThPL model is fainter than the PPL model since there are more, lower-energy, electrons producing unabsorbed synchrotron photons (Figure~\ref{fig:elec_dists}).  Our simulation ends (see discussion in Section~\ref{sec:full_dist}) as the PPL afterglow peaks in radio, but the ThPL radio afterglow is still rising.  It is a reasonable inference that the late-time behavior of the radio afterglow is similar to that of the optical afterglow, discussed below.
  \item The same initial relation between the ThPL and PPL models is visible in the early optical afterglow.  At $\tobs \approx 300$~s, $\nu_{m}$ for the PPL model passes through the optical and the PPL afterglow begins to decline; the thermal peak for the ThPL afterglow does not reach optical frequencies until $\tobs \approx 1600$~s, a factor of 5 later in time.  The ThPL afterglow then experiences a period of steep decline, with $\alpha \approx 1.7$ before optical photons are again produced by the accelerated part of the ThPL distributions.  Once this occurs the ThPL and PPL optical afterglows overlap, as expected from Figures~\ref{fig:p_pl-therm_ssc_compare} and \ref{fig:p_SEDs_ratio}.
  \item The ThPL excess in X-rays is due to the thermal synchrotron peak, and vanishes as the spectral feature passes out of the X-ray waveband.  The same light curve feature appears in the TeV band as well, and at the same time.  The temporal correlation between the light curves in both bands is possibly coincidental, and due to the particular values of the GRB parameters used.  It is nonetheless intriguing, and merits further scrutiny as more GRBs are detected in the TeV band.  The TeV excess persists for longer than the X-ray excess does, since the thermal electrons of the ThPL distributions (being at higher energies than the base of the PPL distributions) can upscatter photons to TeV energies for longer.
  \item The GeV excess is, like the TeV excess, due to thermal-thermal SSC, not to the thermal synchrotron peak.  It occurs later than the TeV excess because the thermal-thermal SSC peak takes time to descend in energy to that waveband. It is also less pronounced than either the X-ray or the TeV excesses, since the spectral feature producing it is already decaying by the time that feature reaches GeV energies (see Figure~\ref{fig:p_pl-therm_ssc_compare}).
\end{itemize}
Finally, note in the bottom panel of Figure~\ref{fig:p_phot_LCs} the thin line labeled ``no SSC''.  This is the TeV light curve when SSC is ignored as an emission process, and it sits a full six orders of magnitude lower than even the PPL TeV light curve.  The fiducial parameters in Table~\ref{tab:GRB_params} are fairly ordinary for GRBs, so the ``no-SSC'' curve highlights that extreme values would be needed to close the gap and allow TeV emission to be produced solely via the synchrotron process.
 
The canonical X-ray afterglow decays with $\alpha \approx 1.2$ \citep{Zhang_etal_2006, Nousek_etal_2006}, and we have drawn a guideline for this decay index in the bottom panel of Figure~\ref{fig:p_phot_LCs}.  Both the X-ray and GeV light curves match this decline, when those photons are coming from the accelerated tail rather than the thermal peak of the electron distributions.  The PPL TeV light curve decays slightly more steeply, with $\alpha \approx 1.4$, and the ThPL TeV light curve would be steeper still since it spends the majority of the observed afterglow decaying towards the PPL values.

Another feature of the canonical afterglow is a shallow plateau before the $\alpha \approx 1.2$ phase.  Figure~\ref{fig:p_phot_LCs} suggests that it is not possible to create this kind of shallow X-ray decay using just thermal particles.  Although the X-ray light curve does have a shallow decay phase (due to the passage of the synchrotron thermal peak), it is joined to the $\alpha \approx 1.2$ phase by a steeper decay rather than by a direct transition.  The work of \citet{ResslerLaskar2017} leads to the same conclusion: a GRB afterglow expanding into a constant-density medium does not have a shallow plateau smoothly connected to the traditional afterglow phase.  A wind-like circumburst medium might allow such an X-ray light curve \citep[see Figure~(5) of][]{GianniosSpitkovsky2009}, but this medium was not considered either here or in \citet{ResslerLaskar2017}.

\begin{figure}
  \epsscale{0.95}
  \includegraphics[width=0.5\columnwidth]{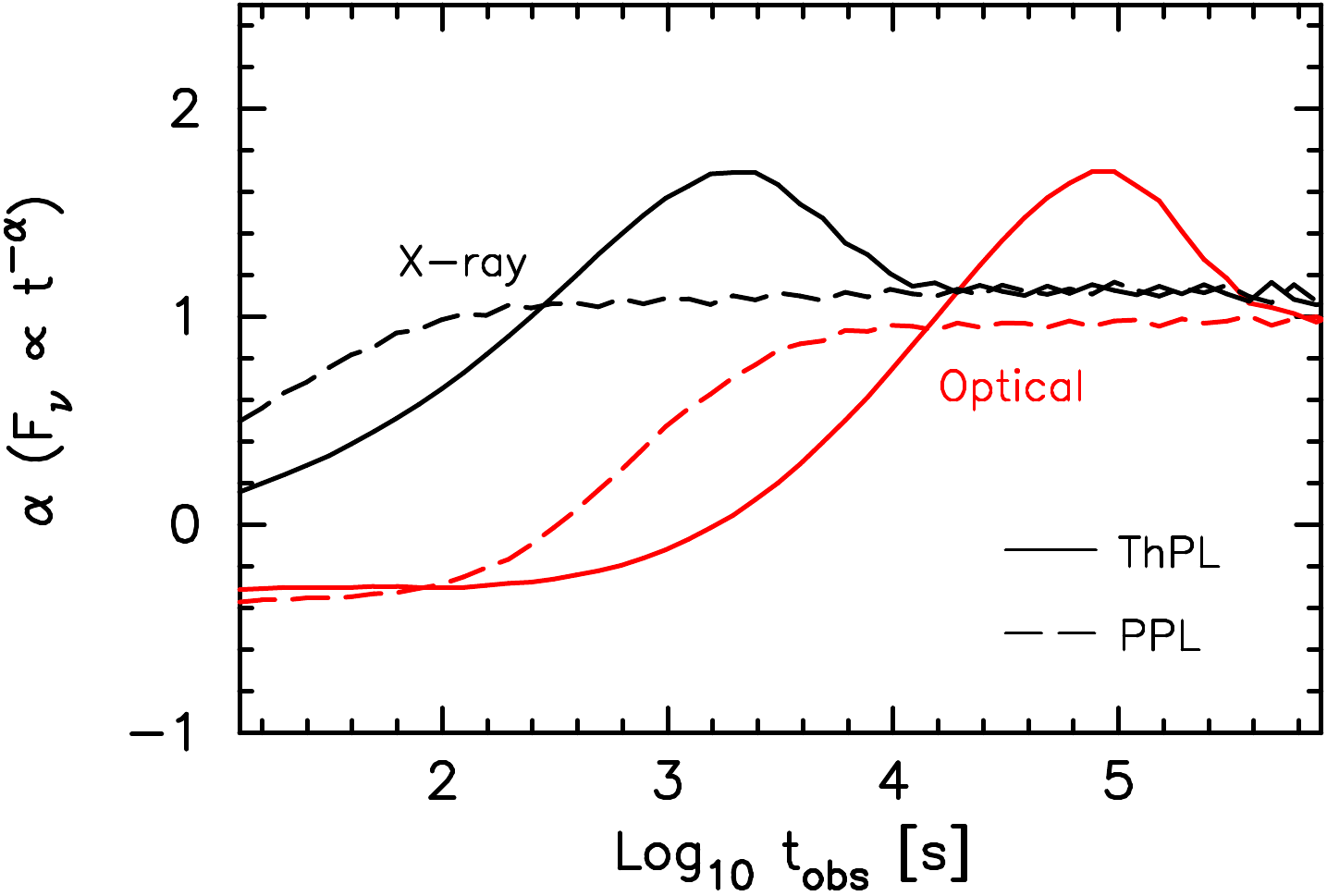}
  \caption{Temporal index $\alpha$ as a function of observer time, in two bands of interest.  Black curves show the evolution in the X-ray band ($0.3-10$~keV), and red curves show optical (640~nm; R band).  Solid lines show the values associated with the ThPL model, and dashed lines are the values from the PPL model.}
  \label{fig:p_Fnu_alphas}
\end{figure}
The X-ray behavior described in the previous paragraph is illustrated in Figure~\ref{fig:p_Fnu_alphas}.  The PPL curve shows the monotonic behavior required to match the canonical X-ray afterglow (although no plateau phase is evident for the observer times plotted); at early observer times the ThPL curve is flatter (smaller $\alpha$) than the PPL curve, after which it steepens to a peak $\alpha \approx 1.7$ and finally relaxes to the PPL value.  The behavior is echoed at optical frequencies, though predictably at later times since the synchrotron peak needs to drop further in energy.  Note, however, that the optical $\alpha$ curves flatten out slightly below the X-ray curves.  This is because the synchrotron cooling break $\nu_{c}$ still lies between the two wavebands at the end of our simulation.  If we ran to later times, $\nu_{c}$ would pass through the optical waveband, and the optical light curve would steepen slightly to match the X-ray light curve \citep[this same behavior was shown in Figure~8 of][]{GianniosSpitkovsky2009}. 

\begin{figure*}
  \epsscale{0.95}
  \includegraphics[height=\textwidth,angle=270]{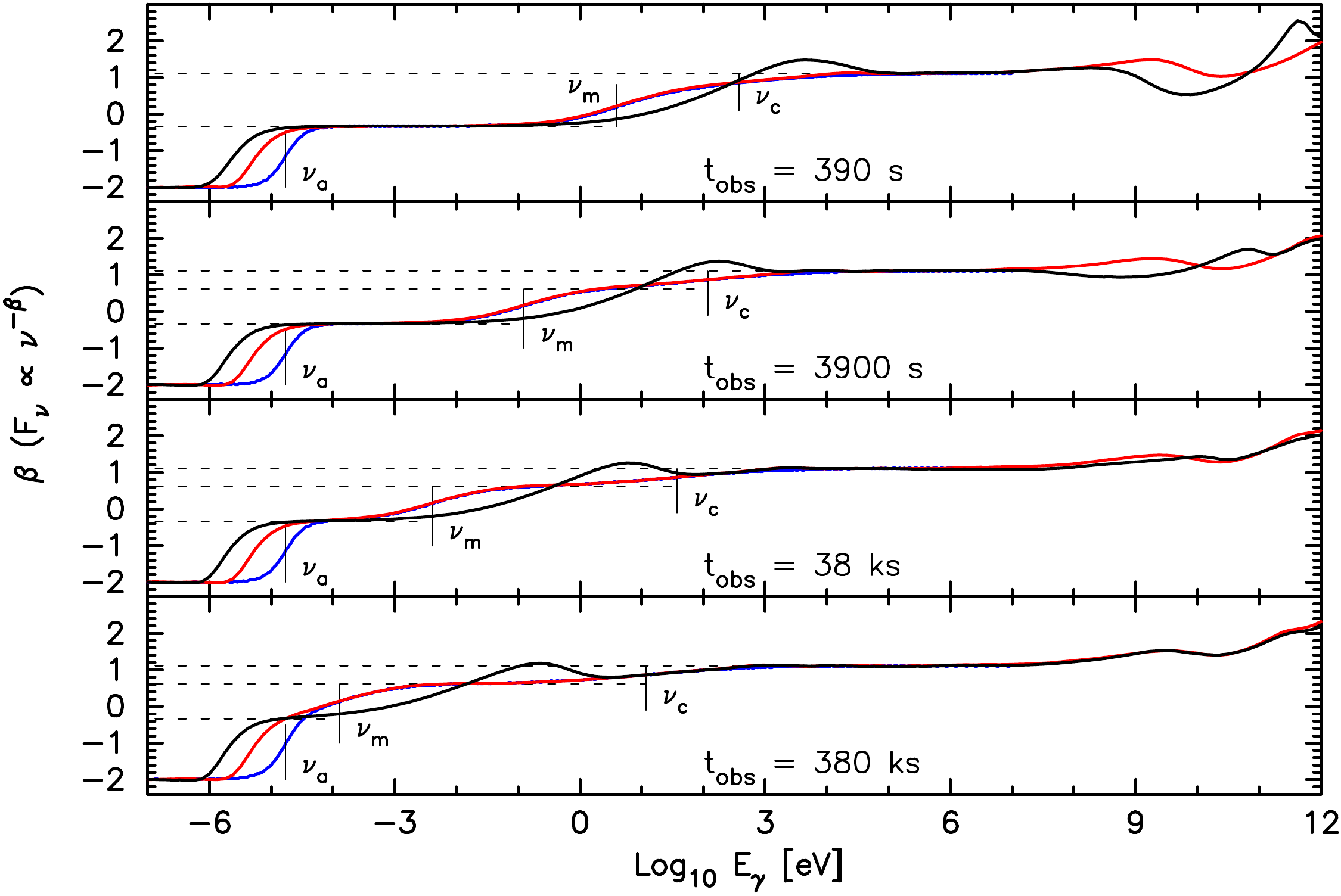}
  \caption{Spectral index of SEDs at four observer times.  Blue curves are based on the fitting formulae of \citet{GranotSari2002}, red curves are based on the PPL model, and black curves are based on the ThPL model.  As in Figure~\ref{fig:p_GS2002_code_check}, we have identified the three spectral breaks associated with a slow-cooling afterglow \citep{GranotSari2002}.  We have also drawn horizontal guide lines for spectral regions associated with the traditional analytical afterglow model.  In the top panel, these are drawn at $F_{\nu} \propto \nu^{1/3}$ and $F_{\nu} \propto \nu^{-p/2}$; in the bottom panels, they are drawn at $F_{\nu} \propto \nu^{1/3}$, $F_{\nu} \propto \nu^{(1-p)/2}$, and $F_{\nu} \propto \nu^{-p/2}$.  Note that the blue curve ends at $E_{\gamma} = 10$~MeV, since the \citet{GranotSari2002} formulae do not include SSC.}
  \label{fig:p_GS2002_saa_betas_wide}
\end{figure*}
We showed broadband SEDs for the afterglow in Figure~\ref{fig:p_SEDs_time_sequence}, but it is also instructive to consider the spectral index $\beta$ as a function of photon energy.  We demonstrate this behavior in Figure~\ref{fig:p_GS2002_saa_betas_wide} for four times during the afterglow.  The break frequencies are marked in order to more easily distinguish spectral regions.

The curves based on the PPL model (red) are in excellent agreement with those computed using the fitting formulae of \citet{GranotSari2002} (blue).  The largest difference is around the SSA break, seen also in Figure~\ref{fig:p_GS2002_code_check} and explained by our use of Equation~\ref{eq:SSA} rather than their Equation~(A20).
The SSA break for the ThPL model (in black) is lower still than that of the PPL model (this was visible in Figure~\ref{fig:p_pl-therm_ssc_compare}, but difficult to discern).  This is a difference from \citep{Warren_etal_2018}; in that previous work, it was implicitly assumed that the NT-only model consisted of electrons in a power law, and a second population (set by the local density) of non-radiating, non-absorbing electrons at much lower energies.

One can see that $\nu_{m} < \nu_{c}$ for all times, consistent with the earlier expectation (Figure~\ref{fig:p_SEDs_time_sequence}) that the afterglow is in the slow-cooling regime.  In the ThPL afterglow, on the other hand, the thermal synchrotron peak is higher in frequency than $\nu_{m}$.  This disparity means the thermal electrons act to delay the transition between fast cooling and slow cooling; indeed, the broad thermal peak is still passing through $\nu_{c}$ at $\tobs \sim 1$~hr.  Once the thermal peak is entirely below $\nu_{c}$, all three models are in in good agreement for their high-energy spectral indices.

The impact of SSC is visible in the top two panels of Figure~\ref{fig:p_GS2002_saa_betas_wide}.  At very early times, thermal-thermal SSC produces a much wider range of $\beta$ than can be achieved with just a power law of electrons: the Comptonized echo of the thermal peak is responsible for the local minimum in the ThPL $\beta$ curve (at $E_{\gamma} = 10$~GeV for $\tobs = 390$~s), and the echo of the exponential rollover in the thermal distribution is responsible for the local maximum in the curve (at $E_{\gamma} \approx 1$~TeV for $\tobs = 390$~s).  At moderate values of $\tobs$ the ThPL and PPL curves have similar values for $\beta$, but at different photon energies.  Given the differences between the early-time curves in Figure~\ref{fig:p_GS2002_saa_betas_wide}, a sufficiently bright burst with good time and energy coverage above GeV provides a good opportunity to discriminate between the PPL and ThPL models.

The GeV-TeV values of $\beta$ for the PPL model do not vary much with time: there is some curvature at a few GeV and a few tens of GeV, but the shape of the spectrum is almost stationary when the emission is produced by a pure power-law distribution.  The ThPL model, on the other hand, sees substantial variation with time as the thermal-thermal SSC peak decays in prominence.  Thus, while both models are capable of producing a TeV $\beta \approx 1.6$ as seen in GRB 180720B \citep{Abdalla_etal_2019Natur575}, only the ThPL model can reproduce the softening in $\beta$ from 1 to 2 that was observed in GRB 190114C \citep{MAGIC2019Natur575}.

\section{Closure relations}
\label{sec:closure_rels}

%
\begin{figure}
  \epsscale{0.95}
  \includegraphics[width=0.5\columnwidth]{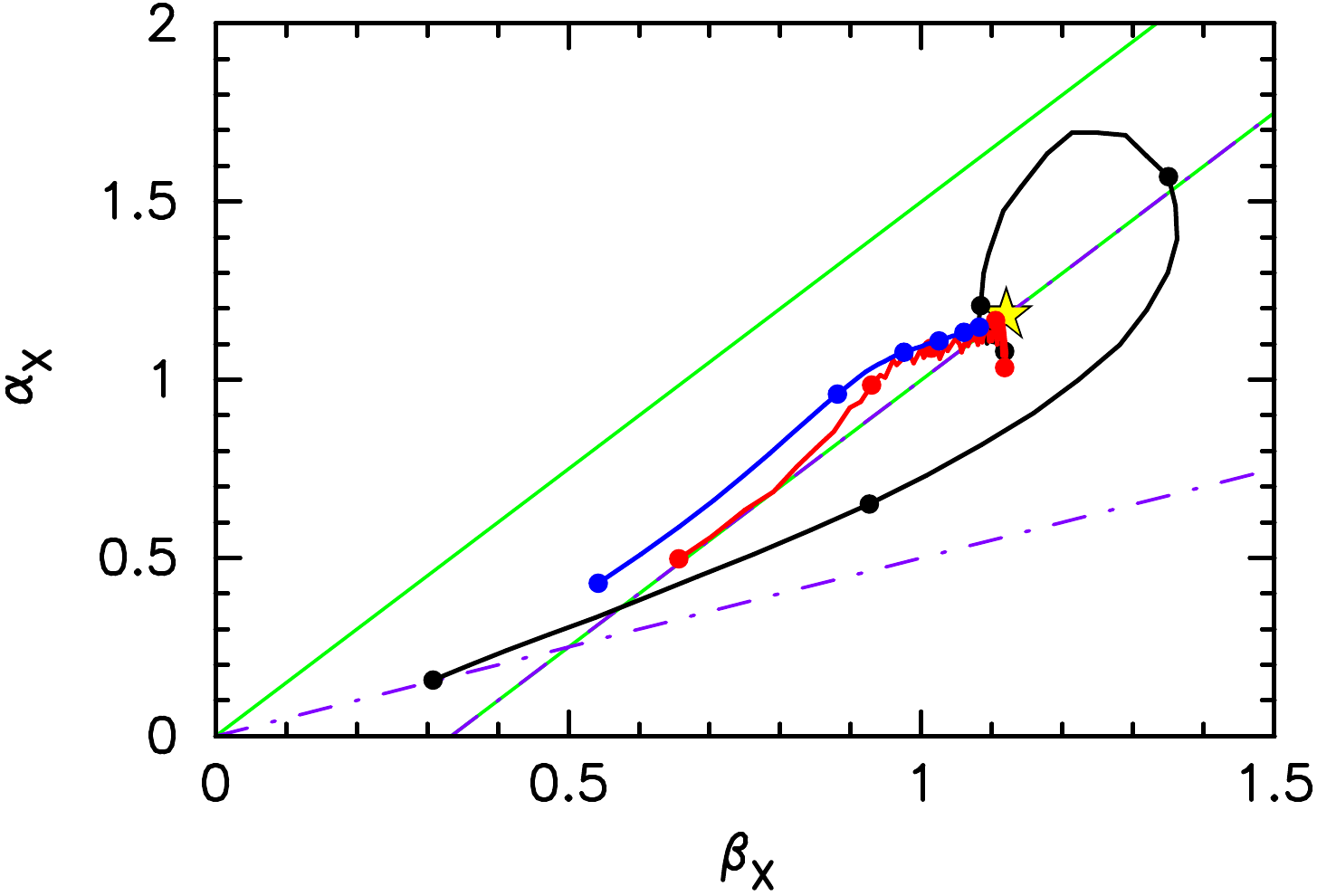}
  \caption{Tracks of $\alpha$ and $\beta$ (in the $0.3-10$~keV energy range) from $\tobs = 10$~s to $\tobs = 10^{6}$~s.  All tracks begin around $\beta_{X} \approx 0.5$ and track up and to the right as time passes.  Black is the ThPL model, red is the PPL model, and blue is derived from the fitting formulae of \citet{GranotSari2002}.  The thin, purple, dash-dotted lines are the predicted closure relations for a GRB afterglow in the fast cooling regime, propagating into a constant-density medium; thin, green, solid lines are the same, but for the slow-cooling regime. Dots mark $\tobs = 10^{1}$~s, $10^{2}$~s, $10^{3}$~s, etc.; the yellow star is the late-time prediction of \citet{GranotSari2002} for a pure power law when SSC is ignored.}
  \label{fig:p_phot_closure-rels}
\end{figure}
Having discussed $\alpha$ and $\beta$ separately in Section~\ref{sec:LCs_indices}, we now restrict our focus to the \textit{Swift} XRT range ($0.3-10$~keV) and plot them both together in Figure~\ref{fig:p_phot_closure-rels}.  This presentation allows for comparison between our models and the closure relations that are frequently employed to interpret X-ray afterglows \citep{SPN1998, ZhangMeszaros2004, Racusin_etal_2009, Srinivasaragavan_etal_2020}.

The four straight lines in Figure~\ref{fig:p_phot_closure-rels} trace the predicted relations between $\alpha$ and $\beta$ for a standard synchrotron afterglow, with $k = 0$ and $p > 2$.  In both the slow-cooling regime (solid green) and the fast-cooling regime (dash-dotted purple), we show a pair of lines.  One green line corresponds to the $\nu_{m} < \nu < \nu_{c}$ part of the SED, and the other line corresponds to the $\nu_{c} < \nu$ part of the SED (the same statements apply to the fast-cooling regime with $\nu_{m}$ and $\nu_{c}$ flipped).  Observed photons must be (1) below the spectral break, (2) above the spectral break, or (3) in the transition region between the two.  Thus the observed $\alpha$ and $\beta$ must fall in the region bounded by the two lines.

The goal in this paper is not to determine which closure relations our afterglows satisfy, in order to diagnose the unknown conditions of an observed GRB.  Instead we are comparing our afterglows to exactly those relations the standard synchrotron model predicts they should satisfy.  We have marked with a star the late-time prediction of Figure~1 in \citet{GranotSari2002}.  There is some early disagreement between the analytical points and the PPL model, caused by the numerical artifact discussed in Figure~\ref{fig:p_GS2002_code_check_hi-res_betas_zoom} of Appendix~\ref{sec:code_check}.  All three models (PPL model, ThPL model, and even the fitting formulae) converge on the same point in $\alpha-\beta$ space, but this point is not the one marked by the star.

While the PPL model and fitting formulae take similar tracks through Figure~\ref{fig:p_phot_closure-rels}, the ThPL model takes a dramatically different path.  Despite exactly the same GRB parameters as the PPL model, the ThPL model spends the early afterglow in the fast-cooling part of the $\alpha-\beta$ space.  As the thermal synchrotron peak (and in particular its exponential turnover before the shock-accelerated tail) passes through the X-ray band, both $\alpha_{X}$ and $\beta_{X}$ increase far beyond the range allowed by a pure power-law electron distribution.  Once the thermal peak is below X-ray energies, those photons are produced by the nonthermal tail of the ThPL distributions, meaning the $\alpha-\beta$ track moves back towards those of the PPL model and fitting formulae.  At very late times, both the PPL and ThPL models are impacted by SSC emission, which causes a slight drop in their $\alpha_{X}$ values.  Given sufficient temporal coverage of a sufficiently bright GRB, it may be possible to produce such tracks and directly test for the shape associated with thermal electrons.

\begin{figure}
  \epsscale{0.95}
  \includegraphics[width=0.5\columnwidth]{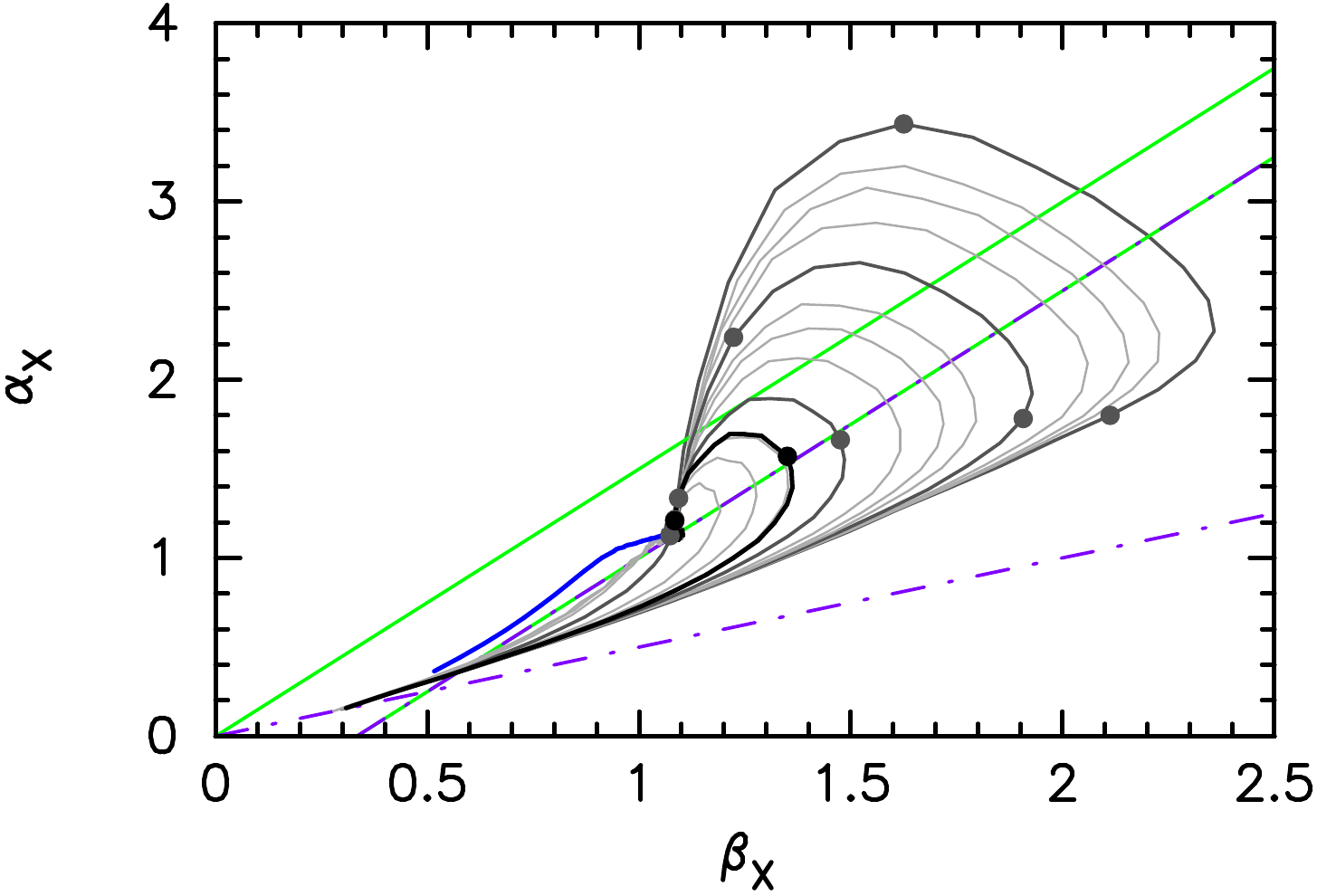}
  \caption{Largely the same as Figure~\ref{fig:p_phot_closure-rels}, but with two changes.  First, time runs from 10~s to $10^{5}$~s since later behavior is not relevant here. Second, we compare only the ThPL model (black and gray curves) to the fitting formulae of \citet{GranotSari2002} (blue curve), omitting the PPL model.  Gray curves show impact of varying $\fNT$ in the ThPL model.  Thicker, dark gray curves show the tracks when $\fNT \in \{10^{-1}, 10^{-2}, 10^{-3}, 10^{-4}\}$; thinner, light gray curves represent $\fNT \in \{2,3,5\} \times \{10^{-1}, 10^{-2}, 10^{-3}, 10^{-4}\}$.  The ThPL track and the dark gray tracks all have two observer times marked; proceeding counterclockwise from the bottom left corner, the first dot is $\tobs = 10^{3}$~s and the second is $\tobs = 10^{4}$~s.}
  \label{fig:p_fNT-test_closure-rels}
\end{figure}
What is the maximum value that $\alpha_{X}$ or $\beta_{X}$ may take?  The peak values in Figure~\ref{fig:p_phot_closure-rels} were produced by the exponential tail of the thermal distribution, so varying $\fNT$---which impacts the height of the thermal peak relative to the power-law tail---should affect the $\alpha-\beta$ track taken by the afterglow. This is indeed the behavior seen in Figure~\ref{fig:p_fNT-test_closure-rels}.
\begin{itemize}
  \item When $\fNT$ is increased from the fiducial value of $1.9\xx{-2}$, the ThPL tracks become more monotonic.  This is sensible, as the electron distribution when $\fNT$ is large has no obvious thermal peak: instead it resembles a smoothly-broken power law with a low-energy index of 2 (see Figure~\ref{fig:p_geometric_fNT} in Appendix~\ref{sec:fNT_max}).  As visible in Figure~\ref{fig:p_fNT-test_closure-rels}, the tracks for large $\fNT$ are monotonic, as is the track associated with the \citet{GranotSari2002} fitting formulae.  There is, however, a slight difference between the tracks of the two models.  The fitting formulae track is ``concave down'', where $\alpha_{X}$ increases more slowly than $\beta_{X}$ does.  ThPL tracks with large $\fNT$ can instead be ``concave up'', where $\beta_{X}$ increases more slowly than does $\alpha_{X}$.  Despite the differences between the tracks, it may be impractical to make observations with enough precision to clearly distinguish the two possibilities should $\fNT$ take large values.
  \item  When $\fNT$ is decreased from the fiducial value, the afterglow tracks have three distinct phases.  In the first phase, both $\alpha_{X}$ and $\beta_{X}$ increase in the fast-cooling part of the space.  The second phase begins when the afterglow reaches its maximum value of $\beta_{X}$, and is marked by decreasing $\beta_{X}$ and increasing $\alpha_{X}$.  Finally, once the peak value of $\alpha_{X}$ is reached, the third phase occurs.  This phase is characterized by a rapid decline in $\alpha_{X}$ accompanied by a slower decline in (or an almost constant) $\beta_{X}$.  The late-time behavior, regardless of the value of $\fNT$, is the same point reached by all afterglows considered in this work.  We wish to emphasize that the afterglow is demonstrably in the slow-cooling regime at all observer times considered (Figure~\ref{fig:p_SEDs_time_sequence}), so these transitions are not due to a change in cooling: they are directly due to the presence of thermal electrons in the distribution.
  \item For $\fNT < 0.01$, the track passes through both the fast-cooling and the slow-cooling regions in the $\alpha-\beta$ plane, and still further into a region that does not satisfy either closure relation.  Observations of GRBs in this part of the space would signal not only the presence of thermal electrons, but that the nonthermal part of the electron distribution is a small fraction of the overall population.
  \item The peak values of both $\alpha_{X}$ and $\beta_{X}$ increase as $\fNT$ decreases.  For exactly the same GRB parameters, inefficient injection into the shock-acceleration process (i.e. a pronounced thermal peak) allows $\alpha_{X}$ to exceed 3, and for $\beta_{X}$ to exceed 2, even though $p = 2.23$ for all the tracks shown in Figure~\ref{fig:p_fNT-test_closure-rels}.
  \item Not only do the extreme values of $\alpha_{X}$ and $\beta_{X}$ increase when $\fNT$ is reduced, but the track evolves more slowly with time.  The two sets of dots in Figure~\ref{fig:p_fNT-test_closure-rels} show the positions of the tracks at observer times of $10^{3}$~s and $10^{4}$~s.  The fiducial model, with $\fNT = 1.9\xx{-2}$, is crossing over from the fast-cooling region to the slow-cooling region at $\tobs = 10^{3}$~s; by $10^{4}$~s it has already reached agreement with the fitting formulae.  In contrast, when $\fNT = 10^{-4}$, the afterglow at $\tobs = 10^{3}$~s has yet to reach its maximum value of $\beta$ (and is still firmly in the fast-cooling part of the figure); and it has only just reached maximum $\alpha$ at $10^{4}$~s.
  \item The afterglows modeled here correspond to Phase III of the canonical afterglow \citep{Zhang_etal_2006, Racusin_etal_2009}.  Figure~2 of \citet{Racusin_etal_2009} presents a histogram of the $\alpha_{X}$ and $\beta_{X}$ values of \textit{Swift} afterglows observed during Phase III.  The values of $\beta_{X}$ cluster around 1.2, with a large but symmetric spread.  The values of $\alpha_{X}$, on the other hand, have a peak around $1.2$ but a significant one-sided tail extending to $\alpha_{X} \approx 3$.  As already mentioned above---and as visible in Figure~\ref{fig:p_fNT-test_closure-rels}---our simulated afterglows show just this behavior at late times.  (While it is possible that selection effects can bias observations of $\alpha_{X}$ and $\beta_{X}$, a plot of GRB fluence versus $\alpha_{X}$ and $\beta_{X}$ showed no clustering of the data, or a particular trend toward bright or faint bursts.  We conclude that there is no clear selection bias causing the distribution of $\alpha_{X}$ and $\beta_{X}$ presented in \citet{Racusin_etal_2009}.)
\end{itemize}

\section{Conclusions}
\label{sec:conclusions}

This work presented a semi-analytical model for GRB afterglows including both thermal electrons and SSC photon production.  Our main results are the following:

\begin{itemize}
  \item We have extended the work of \citet{GranotSari2002} and \citet{ResslerLaskar2017} by including a physically-motivated prescription for both the thermal distribution and the maximum electron energy.  We are also including SSC for the first time in the context of thermal electrons, allowing us to calculate emission across every observable frequency range.
  \item Thermal particles affect the entirety of the afterglow, from GHz to TeV (Figures~\ref{fig:p_pl-therm_ssc_compare}, \ref{fig:p_SEDs_ratio}, and \ref{fig:p_phot_LCs}).  For our fiducial parameters, early-time TeV emission is boosted by more than order of magnitude when thermal particles are included; at late times the ThPL and PPL scenarios predict similar TeV production.  Additionally, X-ray, optical, and radio light curves are all different between the ThPL and PPL scenarios, despite using the same fiducial parameters and the same normalizations for the respective electron distributions (both the overall normalization and that of the accelerated tail).
  \item For the GRB parameters we chose, the Compton $Y$ parameter is much less than 1 for entire simulated afterglow (Figure~\ref{fig:p_compton_Y}).  This parameter is likely to vary locally within the shocked region, and is sure to depend globally on the GRB parameters.  For at least some of the relevant GRB parameter space, though, we retroactively justify our decision to ignore SSC cooling as a contributor to the maximum electron energy.
  \item Thermal electrons impact both temporal decay index $\alpha$ and spectral decay index $\beta$ of emission.  The most basic prediction, and one made previously in the literature \citep{GianniosSpitkovsky2009, WEBN2017}, is non-monotonic evolution of $\alpha$ and $\beta$ with time.  We recover that result in the X-ray band (Figure~\ref{fig:p_Fnu_alphas}), and examination of light curves suggests that a similar behavior happens at any frequency of interest (Figure~\ref{fig:p_phot_LCs}).
  \item In the TeV band specifically, the ThPL model demonstrates changes in $\beta$ with time (Figure~\ref{fig:p_GS2002_saa_betas_wide}).  This is in stark contrast to the PPL model, which predicts no changes in $\beta$, and readily explains observations of TeV afterglows where the spectral index evolved with time.
  \item Closure relations are a common tool for interpreting afterglow observations.  The ThPL and PPL afterglows take very different paths through $\alpha-\beta$ space in X-ray (Figures~\ref{fig:p_phot_closure-rels}, \ref{fig:p_fNT-test_closure-rels}). In particular, if $\fNT < 0.01$ then the ThPL track goes outside both the fast- and slow-cooling closure relations (despite the fact that there is no third possibility in the standard synchrotron afterglow).  Late time X-ray behavior in the ThPL model (with $\fNT < 0.01$)---but not in the PPL model---is consistent with Phase III afterglows as presented in \citet{Racusin_etal_2009}.
  \item As pointed out in previous work, $\fNT$ is a mostly free parameter (though PIC simulations suggest should be in range of $1-3$\%).  Curiously, making the canonical assumption (and one not based on a physical justification) that the power-law part of the ThPL model overlaps with the PPL model causes $\fNT$ to match PIC predictions for the physically relevant range $2.2 \lsim p \lsim 2.5$.  This is, as far as we can tell, a numerical coincidence rather than signaling a deeper physical meaning.
\end{itemize}

Although we have presented here the most complete treatment to date of thermal electrons in GRB afterglows, there are still numerous extensions and refinements that can be made to the model.  An early coasting phase \citep{MLR1993, KPS1999} and energy injection \citep{DaiLu1998, ZhangMeszaros2001} may become observationally relevant if the ratio $E_\mathrm{iso}/n_\mathrm{ext}$ is greater than what was assumed here.  Although we assumed a spherically-symmetric blast wave, all available evidence points to GRB jets having angular structure \citep{LPP2001, RLR2002, Ito_etal_2019, GNB2021}.  Given that jets indeed have angular structure, the viewing angle is likely to lie off the jet axis, breaking the circular symmetry and introducing further complications \citep{Granot_etal_2002, Ryan_etal_2015, Ryan_etal_2020}.  It has also been suggested that a significant portion of observed emission is produced by electrons ahead of the blast wave \citep{SironiSpitkovsky2009ApJL, SKL2015,DerishevPiran2016,DerishevPiran2021}, though the anisotropic nature of the magnetic field there makes computing emission more challenging.  Finally, and perhaps most obviously, the simulated afterglows here assumed a constant-density ambient medium.  Using $k=2$ rather than $k=0$ will affect the hydrodynamic evolution of the blast wave, and in turn the structure and evolution of the SEDs.  Any, or all, of these topics are worth exploring in future papers.

\appendix

\section{Verification of numerical code}
\label{sec:code_check}

%
\begin{figure}
  \epsscale{0.95}
  \includegraphics[width=0.5\columnwidth]{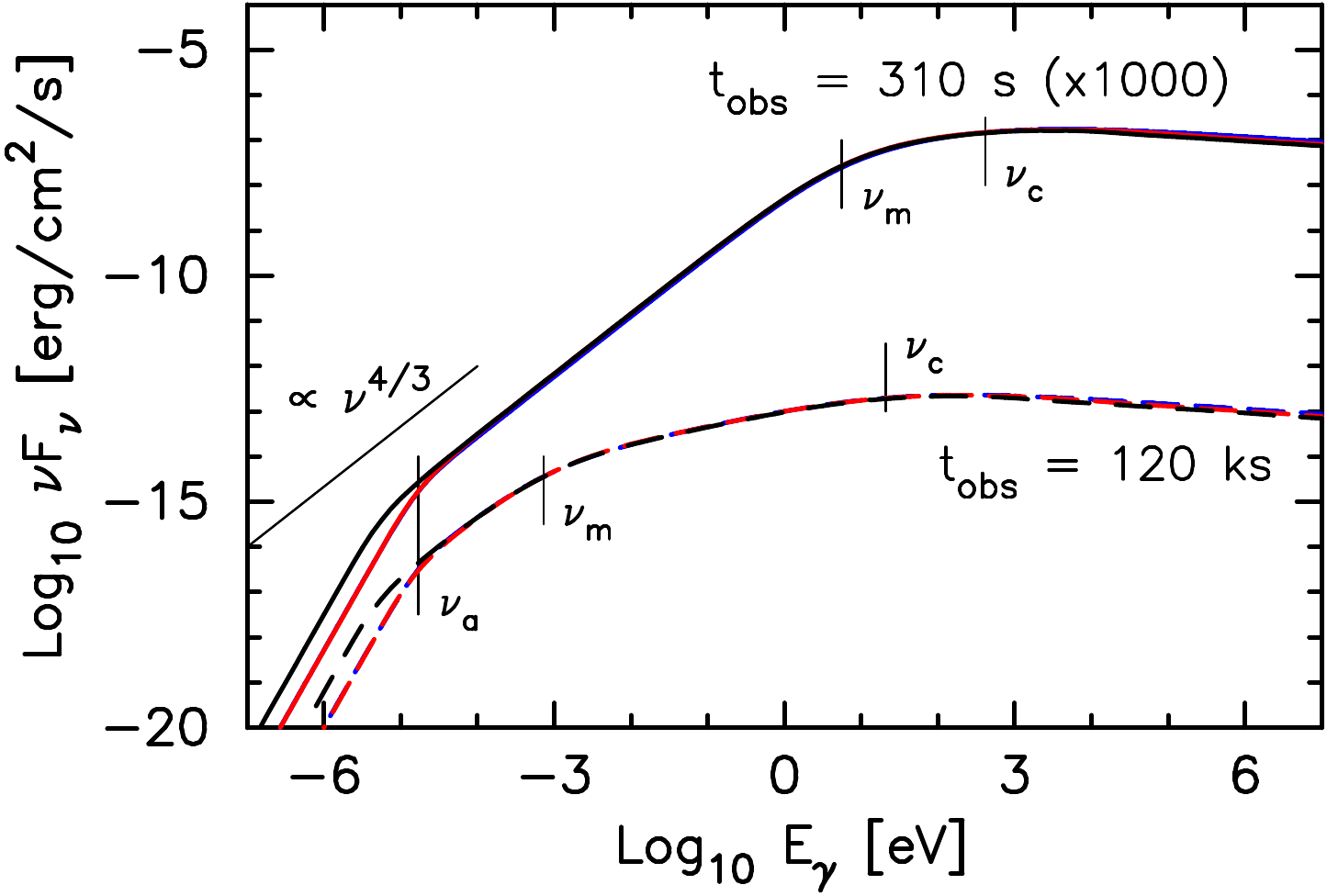}
  \caption{SEDs at two observer times for three different methods of computing afterglow emission.  Solid lines show the SEDs for an observer time of 310~s (scaled by a factor of 1000 for clarity), and dashed lines show the SEDs for an observer time of 120~ks.  Blue lines were computed using Table~2 and Eq.~(5) from \citet{GranotSari2002}; red lines were computed using Eq.~(A24) of that paper; and black lines use the PPL model outlined in Section~\ref{sec:model}.  The three spectral breaks associated with a slow-cooling afterglow (absorption, injection, and cooling) are also identified.}
  \label{fig:p_GS2002_code_check}
\end{figure}
As a test of the procedure outlined in Section~\ref{sec:model}, we compare it against two realizations of the standard synchrotron afterglow.  All three approaches used the GRB parameters listed in Table~\ref{tab:GRB_params}.  In Figure~\ref{fig:p_GS2002_code_check} we show spectral energy distributions (SEDs) for an early and a late observer time.  The blue curves were produced using the fitting formulae of \citet{GranotSari2002}---that is, their Table~2 and Eq.~(5).  The fitting formulae are in good agreement with the SEDs computed using the full radiative transfer integral (Equation~(A24) in that paper), shown in red in Figure~\ref{fig:p_GS2002_code_check}; indeed, the two colors overlap almost perfectly for most of the SEDs shown.  The final pair of curves drawn in Figure~\ref{fig:p_GS2002_code_check} was computed using the PPL model described in Section~\ref{sec:model} (black lines).  To better compare outputs, it was assumed that the non-thermal tail extended to arbitrarily high energies (rather than ending where radiative losses exceed acceleration gains, as in Equation~\ref{eq:elec_max_en}).

For ease of comprehension, we have identified the key break frequencies in Figure~\ref{fig:p_GS2002_code_check}.  Since $\nu_{m} < \nu_{c}$ for both times shown, the afterglow is in the slow-cooling regime.  As pointed out in Equation~\ref{eq:t_fastslow}, the transition from fast to slow cooling happens extremely early for the GRB parameters used, so this ordering of break frequencies is to be expected.

The three computing methods agree quite well for most of the energy range shown in Figure~\ref{fig:p_GS2002_code_check}.  The most obvious difference is at the low end of the SEDs: specifically, the location of $\nu_{a}$, at which the afterglow becomes optically thick.  The black curve in Figure~\ref{fig:p_GS2002_code_check} predicts an absorption frequency notably lower than do the two methods presented in \citet{GranotSari2002}, due to our use of Equation~\ref{eq:SSA} rather than Equation~(A20) in the latter paper.

\begin{figure*}
  \epsscale{0.95}
  \includegraphics[height=\textwidth,angle=270]{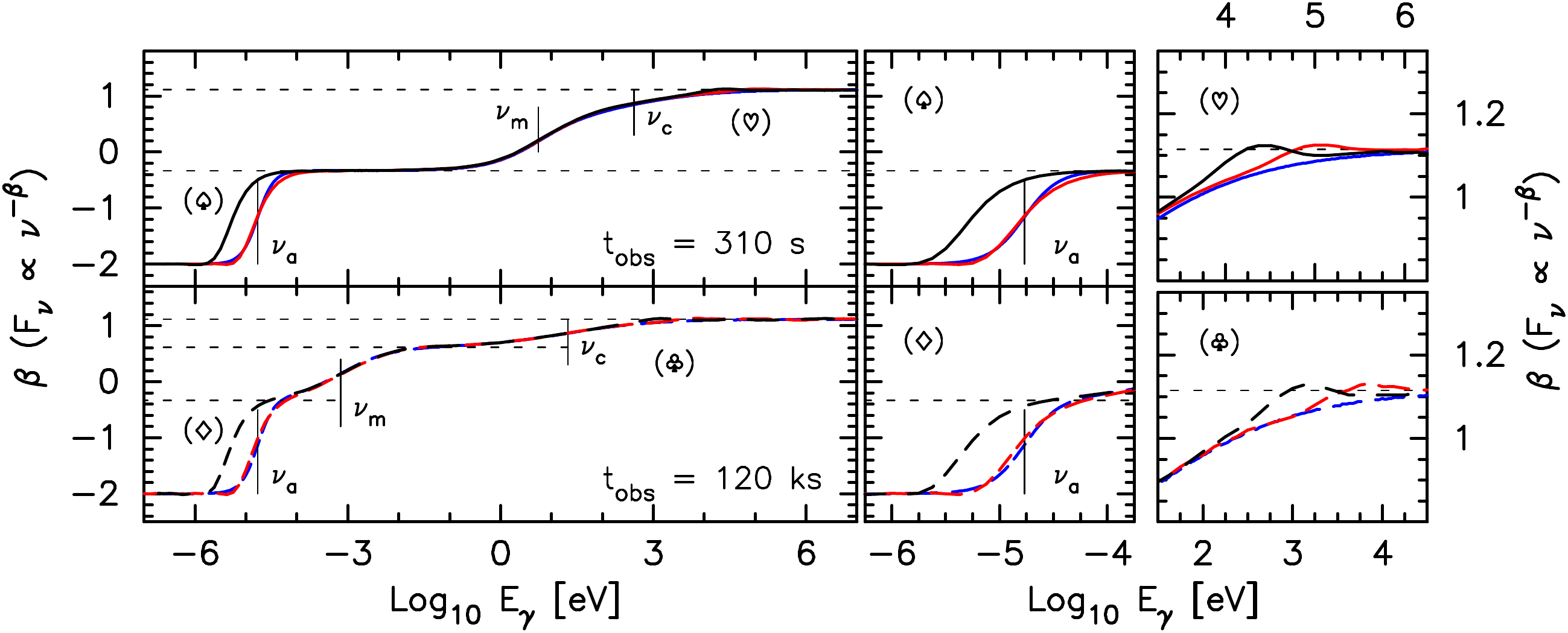}
  \caption{Spectral index of SEDs at two observer times.  Colors and line types are as in Figure~\ref{fig:p_GS2002_code_check}.  As in Figure~\ref{fig:p_GS2002_code_check}, we have identified the three spectral breaks associated with a slow-cooling afterglow in \citet{GranotSari2002}.  We have also drawn horizontal guide lines for spectral regions associated with the traditional analytical afterglow model.  In the top panel, these are drawn at $F_{\nu} \propto \nu^{1/3}$ and $F_{\nu} \propto \nu^{-p/2}$; in the bottom panel, they are drawn at $F_{\nu} \propto \nu^{1/3}$, $F_{\nu} \propto \nu^{(1-p)/2}$, and $F_{\nu} \propto \nu^{-p/2}$. In the left two panels, we have identified four locations with the symbols $\varspadesuit$, $\heartsuit$, $\diamondsuit$, and $\varclubsuit$.  The right four panels expand the region identified by each symbol, to more clearly show the differences between each set of curves.}  
  \label{fig:p_GS2002_code_check_hi-res_betas_zoom}
\end{figure*}

Beyond the significant difference at low energies, there is also slight disagreement at high photon energies---on the order of $10-15\%$---which persists regardless of the integration method implemented or the fineness of the various discretizations used  (photon energy, angular integration, radiative transfer integration). Inspection of Figure~\ref{fig:p_GS2002_code_check_hi-res_betas_zoom} reveals the reason for this discrepancy.  This figure shows the spectral index $\beta$ as a function of photon energy for the same energy range as Figure~\ref{fig:p_GS2002_code_check}.  It can be seen that the blue curves \citep[based on the fitting formulae in][]{GranotSari2002} smoothly vary in the vicinity of $\nu_{c}$, as indeed they are mathematically required to do.  As the cooling break transitions into the highest-energy power-law segment \citep[PLS H in Figure~1 of][]{GranotSari2002}, the two numerical integrators (in red and black) behave differently.  Rather than a smooth transition to the eventual value of $\beta$, they continue to increase more steeply, and even slightly overshoot the endpoint before relaxing back to it.  This larger value of $\beta$ means their SEDs decline more rapidly than the SED of the fitting formulae, which causes the slight disparity at high photon energies in Figure~\ref{fig:p_GS2002_code_check}.

The key result from Figures~\ref{fig:p_GS2002_code_check} and \ref{fig:p_GS2002_code_check_hi-res_betas_zoom} is that our code correctly reproduces the key spectral features and temporal behavior of the fitting formulae in \citet{GranotSari2002}.  The slight discrepancy around $\nu_{c}$ appears to be numerical in nature rather than physical.  As it seems to behave in a consistent manner regardless of observer time, it could be corrected during or after the calculation should percent-level accuracy ever become necessary.

\section{The synchrotron absorption coefficient $\alpha_{\nu}$}
\label{sec:alpha_nu}

In order to compute synchrotron self-absorption, we employed Equation~(6.50) of \citet{RybickiLightman1979} (Equation~\ref{eq:SSA} here).  
A different form of the equation is used in \citet{GranotSari2002}.  In a footnote in that paper, those authors state that ``eq.~(18) of \citet{GPS1999ApJ527}, which is essentially eq.~(6.52) of \citet{RybickiLightman1979}, misses the term associated with the discontinuity at the lower edge of the electron distribution (at $\gamma_\mathrm{min}$) when derived from eq.~(6.50) of Rybicki \& Lightman.  This missing term caused an overestimation of the absorption coefficient by a factor of $f = 3(p+2)/4$''.

The alternate equation is found in Appendix~A of that paper:
\begin{linenomath}\begin{equation}
  \alpha_{\nu,\mathrm{GS}} = \frac{ 1 }{ 8 \pi m \nu^{2} } \int_{\gamma_\mathrm{min}}^{\gamma_\mathrm{max}} \frac{ N(\gamma) }{ \gamma^{2} } \, \frac{ \partial }{ \partial\gamma } \left[ \gamma^{2} P(\nu,\gamma) \right] d\gamma
  \label{eq:GS2002_A20}
\end{equation}\end{linenomath}
Note that the equation involves the derivative of $P(\nu,\gamma)$ rather than of $N(\gamma)$ $(= dn_{\gamma}/d\gamma)$ as in Equation~(6.50) of \citet{RybickiLightman1979}. The two equations are linked through integration by parts:
\begin{linenomath}\begin{align}
  \alpha_{\nu,\mathrm{GS}} &= -\frac{ 1 }{ 8 \pi m \nu^{2} } \int_{\gamma_\mathrm{min}}^{\gamma_\mathrm{max}} \gamma^{2} P(\nu,\gamma) \frac{ \partial }{ \partial\gamma } \left[ \frac{ N(\gamma) }{ \gamma^{2} } \right] d\gamma + \frac{ 1 }{ 8 \pi m \nu^{2} } \left[ N(\gamma) P(\nu,\gamma) \right]_{\gamma_\mathrm{min}}^{\gamma_\mathrm{max}}
  \label{eq:GS2002_int_by_parts}
\end{align}\end{linenomath}
The first term in Equation~\ref{eq:GS2002_int_by_parts} is simply Equation~\ref{eq:SSA}, while the second term contains information about the endpoints of the distribution.

Let us assume that $\gamma_\mathrm{max} \rightarrow \infty$, and that we are dealing with a power-law distribution of electrons, $N(\gamma) = K_\mathrm{pl} \gamma^{-p}$, whose index $p > 2$.  Then the $\gamma_\mathrm{max}$ endpoint term vanishes (since $N(\gamma_\mathrm{max}) \rightarrow 0$), and Equation~\ref{eq:GS2002_int_by_parts} may be rewritten
\begin{linenomath}\begin{align}
 \alpha_{\nu,\mathrm{GS}} &= \Omega \, n \, \frac{ 3(p-1)(p+2) }{ 3p+2 } \, \gamma_\mathrm{min}^{-5/3} - \Omega \, K_\mathrm{pl} \, \gamma_\mathrm{min}^{-p - \frac{2}{3}} \nonumber \\
  &= \Omega \, n \, \frac{ 3(p-1)(p+2) }{ 3p+2 } \, \gamma_\mathrm{min}^{-5/3} - \Omega \, n \, (p-1) \, \gamma_\mathrm{min}^{-5/3} \nonumber \\
  &= \Omega \, n \, \frac{ 4(p-1) }{ 3p+2 } \, \gamma_\mathrm{min}^{-5/3} .
  \label{eq:GS2002_alphanu}
\end{align}\end{linenomath}
In the above equation we have assumed that $P(\nu,\gamma)$ lies in the $\nu^{1/3}$ part of the spectrum, and we have collected numerical prefactors into the single term $\Omega$ for easier readability; the substitution $K_\mathrm{pl} = n (p-1) \gamma_\mathrm{min}^{p-1}$ comes from the normalization condition $\int N d\gamma = n$ for a desired number density $n$.

Repeating the above exercise for Equation~\ref{eq:SSA} leads to \citep{Warren_etal_2018}
\begin{linenomath}\begin{equation}
  \alpha_{\nu,\mathrm{RL}} = \Omega \, n \, \frac{ 3(p-1)(p+2) }{ 3p+2 } \gamma_\mathrm{min}^{-5/3} ,
\end{equation}\end{linenomath}
which is greater by a factor $3(p+2)/4$ than the end result of Equation~\ref{eq:GS2002_alphanu}.  We therefore use Equation~\ref{eq:SSA} to compute the absorption coefficient rather than Equation~\ref{eq:GS2002_A20}.

\section{Maximum $\fNT$ with a ``thermal'' distribution}
\label{sec:fNT_max}

%
\begin{figure}
  \epsscale{0.95}
  \includegraphics[width=0.5\columnwidth]{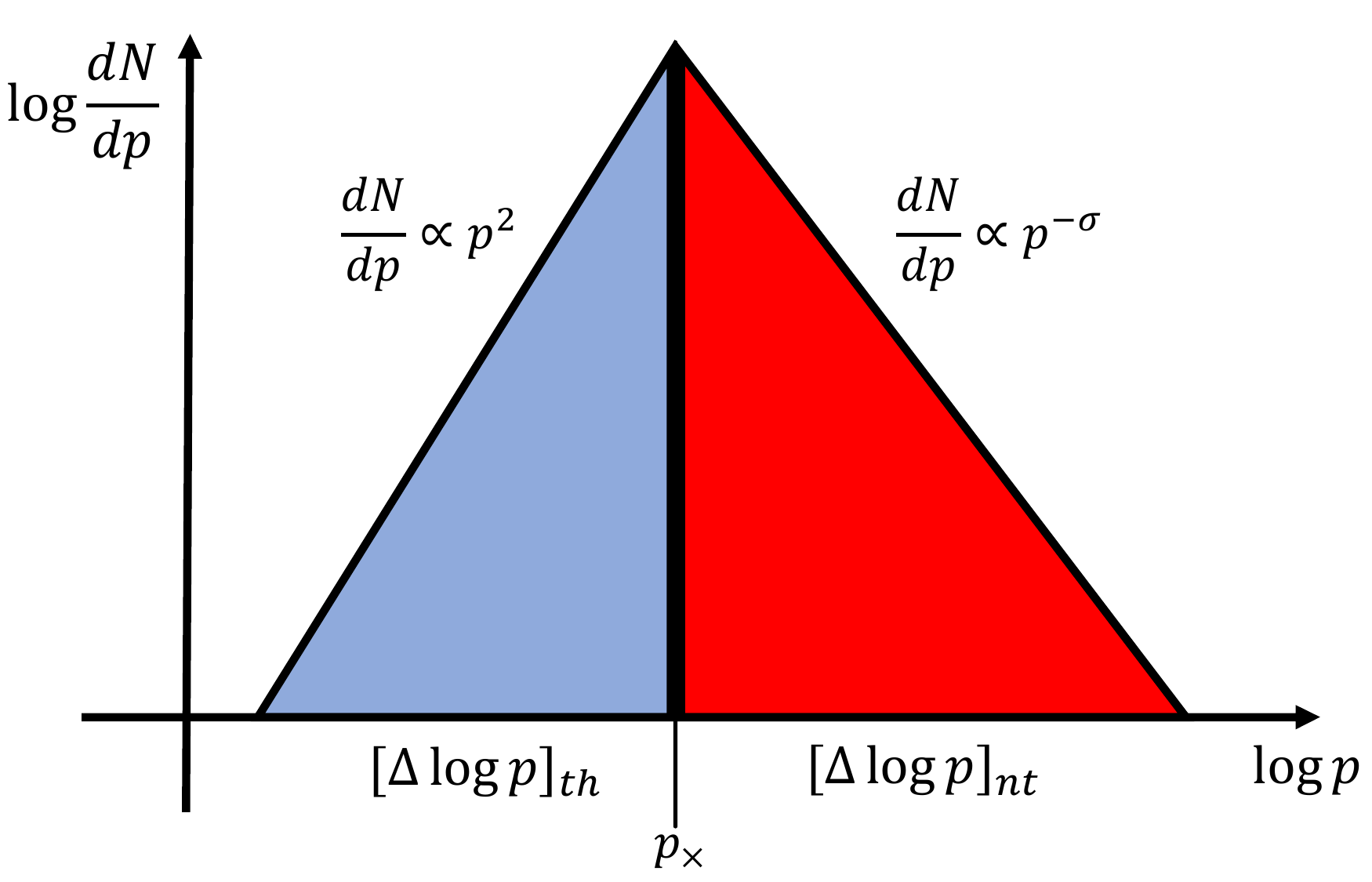}
  \caption{Illustration of a scenario leading to the maximum possible value of $\fNT$.  Note that we use momentum (rather than energy or Lorentz factor) as our dependent variable.  To avoid using the same symbol twice for different quantities, we therefore refer to the non-thermal spectral index as $\sigma$ here rather than as $p$ (which it is called elsewhere in this paper).}
  \label{fig:p_geometric_fNT}
\end{figure}
For a thermal distribution, the maximum value of $\fNT$ occurs when $p_{\times}$ lies below the thermal peak completely, falling instead in the $p^{2}$ low-energy part of the distribution.  Such a situation is illustrated in Figure~\ref{fig:p_geometric_fNT}, but note the different use of symbols here as opposed to the rest of this work: we use $p$ to denote electron momentum rather than a spectral index, and we instead use $\sigma$ for the spectral index of the accelerated electron population.

The thermal particle number can be calculated as
\begin{linenomath}\begin{equation}
    N_{th} = \int_0^{p_\times} C p^2 dp = C \frac{ p_\times^3 }{ 3 }
\end{equation}\end{linenomath}
where $C$ is a normalization constant. The number of non-thermal particles is
\begin{linenomath}\begin{equation}
    N_{nt} = \int_{p_\times}^\infty C p_\times^{2+\sigma} p^{-\sigma} dp = C \frac{p_\times^3}{\sigma -1}
\end{equation}\end{linenomath}
with a normalization constant $C p_{\times}^{2+\sigma}$ required so the two parts of the distribution meet at $p_{\times}$.  Thus
\begin{linenomath}\begin{align}
  f_{\mathrm{NT},\mathrm{max}} &= \frac{ N_{nt}  }{  N_{nt} +  N_{th} } \nonumber \\
  &= \frac{ 3 }{ \sigma + 2 } ,
\end{align}\end{linenomath}
precisely the formula recovered from Equation~(17) of \citet{ResslerLaskar2017}.

\acknowledgments 
This work was funded in part by the Interdisciplinary Theoretical and Mathematical Sciences (iTHEMS, \url{https://ithems.riken.jp}) program at RIKEN (DCW, SN).  MB would like to acknowledge support by NASA grants 80NSSC17K0757 and NNH19ZDA001N-FERMI, and NSF grants 10001562 and 10001521. This work is supported by JSPS Grants-in-Aid for Scientific Research “KAKENHI” (A: Grant Number JP19H00693).  SN also acknowledges the support from Pioneering Program of RIKEN for Evolution of Matter in the Universe (r-EMU). BA acknowledges support from the Swedish Research Council (grant reference number 2020-00540). HI acknowledges support from  JSPS KAKENHI Grant Number JP19K03878 and JP20H04751.

\bibliographystyle{aa} 
\bibliography{dcw}

\end{document}